\documentclass{article}

\usepackage{PRIMEarxiv}

\usepackage[utf8]{inputenc} 
\DeclareUnicodeCharacter{202F}{\,}
\usepackage{float}
\usepackage[T1]{fontenc}    
\usepackage{hyperref}       
\usepackage{url}            
\usepackage{booktabs}       
\usepackage{amsfonts}       
\usepackage{nicefrac}       
\usepackage{microtype}      
\usepackage{lipsum}
\usepackage{fancyhdr}       
\usepackage{graphicx}       

\usepackage{longtable}
\usepackage{tabularx}
\usepackage{multirow}
\usepackage{multicol}
\usepackage{tabularray}
\usepackage{array}
\usepackage{caption}

\usepackage{hyperref}
\usepackage{subfigure}

\title{Deep Dive into MRI: Exploring Deep Learning Applications in 0.55T and 7T MRI}
\author{
    Ana Carolina Alves\textsuperscript{1,2},
    André Ferreira\textsuperscript{1,2,3},
    \textbf{Behrus Puladi\textsuperscript{3,5}},\\ 
    \textbf{Jan Egger\textsuperscript{2,4,6}},
    \textbf{Victor Alves\textsuperscript{1}}
}

\begin{document}

\maketitle
\textsuperscript{1} Center Algoritmi / LASI, University of Minho, Braga, Portugal

\textsuperscript{2} Institute for Artificial Intelligence in Medicine, University Medicine Essen, Essen, Germany

\textsuperscript{3} Department of Oral and Maxillofacial Surgery, University Hospital RWTH Aachen, Aachen, Germany

\textsuperscript{4} Institute of Computer Graphics and Vision (ICG), Graz University of Technology, Graz, Austria

\textsuperscript{5} Institute of Medical Informatics, University Hospital RWTH Aachen, Aachen, Germany

\textsuperscript{6} Center for Virtual and Extended Reality in Medicine, University Medicine Essen, Essen, Germany

\paragraph{}
\paragraph{}
\paragraph{}

\begin{abstract}
The development of magnetic resonance imaging (MRI) for medical imaging has provided a leap forward in diagnosis, providing a safe, non-invasive alternative to techniques involving ionising radiation exposure for diagnostic purposes.
It was described by Block and Purcel in 1946, and it was not until 1980 that the first clinical application of MRI became available.
Since that time the MRI has gone through many advances and has altered the way diagnosing procedures are performed. Due to its ability to improve constantly, MRI has become a commonly used practice among several specialisations in medicine. Particularly starting 0.55T and 7T MRI technologies have pointed out enhanced preservation of image detail and advanced tissue characterisation.
This review examines the integration of deep learning (DL) techniques into these MRI modalities, disseminating and exploring the study applications. It highlights how DL contributes to 0.55T and 7T MRI data, showcasing the potential of DL in improving and refining these technologies.
The review ends with a brief overview of how MRI technology will evolve in the coming years.
\end{abstract}


\paragraph{}
\textbf{\emph{Abbreviations:}}
ADC: apparent diffusion coefficient; ADLR: advanced DL-based reconstruction; ALS: amyotrophic lateral sclerosis; bSSFP: balanced steady state free precession; CMR: cardiovascular magnetic resonance; CNN: convolutional neural network; CNR: contrast-to-noise ratio; DBS: deep brain stimulation; DL: deep learning; DWI: diffusion-weighted imaging; ECG: electrocardiogram; EPI: echoplanar imaging; FDA: food and drug administration; FIST: fast imaging with steady-state precession; HD: huntington disease; IAC: internal auditory canal; LWD: lung water density; MRF: magnetic resonance fingerprinting; MRI: magnetic resonance imaging; RF: radiofrequency; MS: multiple sclerosis; MSA: Multiple System Atrophy; MSK: musculoskeletal; MWF: myelin water fraction; MWI: myelin water image; NEXs: number of excitations; PD: parkinson disease; PJI: periprosthetic joint infection; PTx: parallel transmit; PVS: perivascular spaces; qRIM: quantitative Recurrent Inference Machine; QSM: quantitative susceptibility mapping; RDN: residual dense network; SLED: self-labelled encoder-decoder; SNR: signal-to-noise ratio; THA: total hip arthroplasty; TMDs: temporomandibular disorders; TMJ: temporomandibular joint; TSE: turbo spin-echo; UHF: ultra high field; UTE: ultrashort echo time; VS: vestibular schwannomas.
\section{Introduction}
\paragraph{}
Since it was first introduced in the early 1980s, magnetic resonance imaging (MRI) has become an essential non-invasive diagnostic tool in modern healthcare. This imaging modality has been improved, making it better suited to different medical specialities. The most recent advances are the 0.55T (Tesla) and 7T MRI systems. The introduction of these systems marks the continued evolution of MRI technology, offering better image resolution, faster imaging processes and more accurate diagnoses, which means MRI can be used in a wider range of clinical applications \cite{Grover2015}.
The application of artificial intelligence and deep learning (DL) processing MRI data, belongs to the most recent stages of MRI development and represents a notable improvement as it opens up new possibilities to improve image quality, reduce the time required to obtain images and improve the diagnostic ability of MRI. 
DL is an artificial intelligence technique that is effective in identifying complex patterns in large data sets to help doctors make decisions. It is being increasingly applied to MRI to address inherent limitations in image speed and resolution \cite{Hossain2024}.
\paragraph{}
This comprehensive review examines the cutting-edge advances and clinical implementations of DL for MRI, focused on the 0.55T and 7T field strengths. Given the rapid pace at which this field is evolving, it is challenging to summarise all of the published literature. Previous reviews have identified the fundamental components of DL architectures and MRI technologies at different magnetic field strengths \cite{Grover2015}, \cite{McMahon2011}, \cite{Hossain2024}, \cite{Tian_review}, \cite{Pogarell2024}, \cite{Arnold2023}, \cite{Barisano2019}, \cite{Shaffer2022}, \cite{Vachha2021}.
This review specifically explores how emerging 0.55T, 7T, and DL methods can synergistically elevate medical imaging capabilities and patient care standards.
This review was conducted according to the PRISMA diagram \cite{Prisma}. The methodological approach involved a comprehensive analysis of high-field and low-field MRI systems, including research papers on DL. The following databases were searched: PubMed, Science Direct, and Scopus, using two search queries '("0.55T MRI" AND "deep learning")' and '("7T MRI" AND "deep learning")' to identify specific papers on the different topics mentioned above. 
\paragraph{}
The initial search query was employed in Section 2 for a review of the applications of DL in 0.55T MRI. The search yielded 11 articles, two from PubMed, seven from Science Direct, and two from Scopus. Following the removal of duplicates (4) and ineligible records (5), two articles were identified for review. The following records were considered to be ineligible for inclusion in the study: encyclopedias, book chapters, conference abstracts, and editorials. 
A total of two articles were identified for inclusion following the initial search query. Nevertheless, to gain a more profound comprehension of low-field strength MRI, an additional search query was implemented: '("0.55T MRI")'. The additional search yielded an additional 47 references, bringing the total number of references included in this manuscript to 49. The 49 references included in this manuscript are composed of seven review articles and two website links.
\paragraph{}
The second search query was employed in Section 3 for a review of the applications of DL in 7T MRI. The initial search yielded 688 articles, 64 from PubMed, 577 from Science Direct, and 47 from Scopus. Following the removal of duplicates (205), ineligible records (322), and records not in the English language (9), 152 articles were identified for examination. Following the screening of the titles and abstracts, an additional 99 articles were excluded from the review process. The exclusion of these articles was based on the absence of information in the abstracts related to 7T or other related topics. A total of 53 articles were identified as suitable for inclusion in this manuscript. The final selection includes four review articles.

\section{Fundamentals of 0.55T and 7T MRI}
\paragraph{}
MRI is a diagnostic modality that stands out by not using ionising radiation, and at imaging soft tissues, such as the brain and muscles. An MRI system uses the magnetic properties of water and fat protons to generate highly detailed 3D anatomical images \cite{Grover2015}. This section outlines the fundamental principles of MRI technology, as well as the characteristics and developments of the 0.55T and 7T MRI systems. 

\subsection{Principles of MRI Technology}
\begin{figure}[ht]
\centering
\includegraphics[height=6cm]{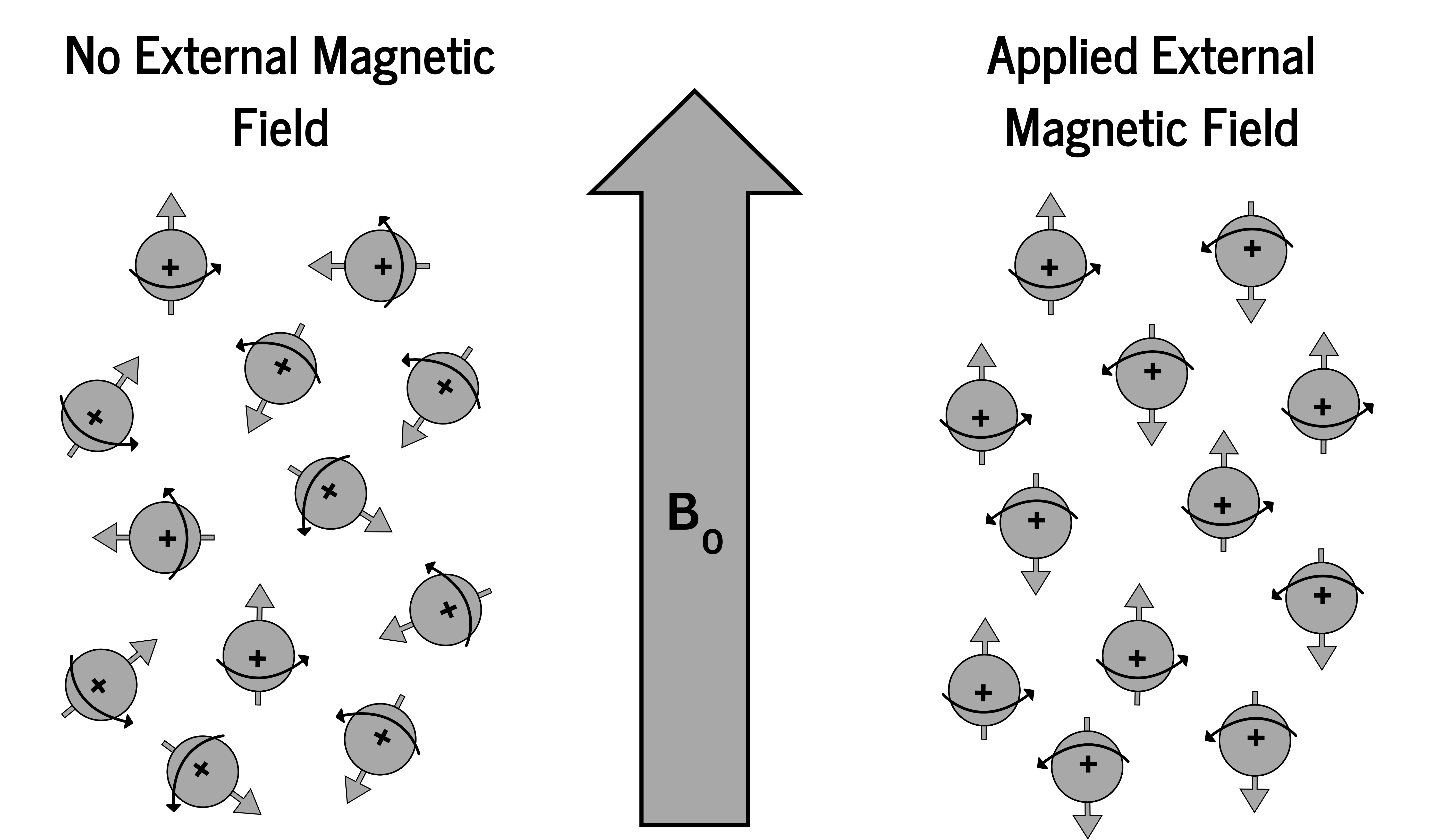}
\caption{Without an external magnetic field, the proton's spin axes are oriented randomly in various directions. However, when a magnetic field is applied, more protons tend to align their spin axes along the direction of the applied magnetic field, creating an excess of protons whose spin orientation matches the field direction.}
\label{fig1}
\end{figure}

\paragraph{}
MRI operates by using an intense magnetic field that aligns the protons in the tissues of a person. The spins of protons align parallel to it when it is present. The protons are displaced momentarily from their normal position when they receive an RF pulse. After the completion of the RF pulse, the protons again realign with the magnetic field and emit energy \cite{McMahon2011}. This emitted energy is captured by MRI sensors and used to create images which represent body structures as shown in Figure \ref{fig1}.
\paragraph{}
MRI has some advantages over other imaging modalities. It does not use ionising radiation, making it safe for patients especially those who have to undergo frequent scans. Nevertheless, strong magnetic fields used in MRI may cause problems for certain metallic objects thereby posing risks to people who have specific implants \cite{Grover2015}.
Thus, it is necessary to screen patients for metal implants, consider possible effects on pregnant patients, handle claustrophobia and cope with concerns about noise and nerve stimulation during MRI scans \cite{McMahon2011}.
\subsection{Evolution of 0.55T and 7T MRI Systems}
\paragraph{}
The magnetic field intensity of MRI systems influences their performance by changing the signal-to-noise ratio (SNR), which in turn affects the quality of a resultant image. 
According to their field strength, MRI systems are rated as low-field, high-field or ultra-high-field systems.
\paragraph{}
\textbf{Low-Field MRI - 0.55T}
\paragraph{}
Low-field MRI systems, like the 0.55T MRI, function at a decreased magnetic field strength, leading to reduced SNR and CNR. This often leads to a lowering of spatial resolution and a longer duration for scanning. 
However, advancements in technology have improved the capabilities of low-field MRI. Advanced hardware and sophisticated image reconstruction techniques greatly improve the diagnostic capabilities of modern 0.55T MRI systems \cite{Pogarell2024}, \cite{Tian_review}.
Low-field MRI's open design provides a major benefit by enhancing patient accessibility and comfort, particularly for individuals with claustrophobia. Moreover, low-field MRI systems are usually more cost-effective, needing minimal shielding and consuming less power, ultimately cutting down operational expenses \cite{Arnold2023}.
\paragraph{}
\textbf{High-Field and Ultra-High-Field MRI - 7T and above}
\paragraph{}
To achieve enhanced image resolution and a more rapid scan time, physicians use MRI systems with high magnetic fields, such as 1.5T and 3T \cite{Barisano2019}. At the same time, ultra-high field MRI systems that focus mainly on 7T and above are leading the way in terms of MRI technology. 
Consequently, the first 7T clinical MRI scanner was launched in 2017, marking a significant advance in this technology that is now applicable at both clinical and research levels \cite{Shaffer2022}, \cite{Vachha2021}.
This results in an increase in spatial resolution and SNR, allowing the use of smaller voxels and the presentation of more detailed anatomical structures \cite{Barisano2019}. 
This proves advantageous, especially when dealing with neurological diseases, along with musculoskeletal and cardiovascular images.
However, UHF MRI has its limitations. The higher intensity magnet can cause dizziness or nausea, as well as artefacts in the resulting images. In addition, processing and storing the large amounts of data produced requires considerable computing power \cite{Barisano2019}, \cite{Shaffer2022}.
\newpage
\paragraph{}
\textbf{Integration of Deep Learning with MRI}
\paragraph{}
Incorporating DL methods into MRI technology is a breakthrough, providing new methods to enhance image quality, speed up imaging processes, and enhance diagnostic precision. 
DL algorithms are being used more in MRI to address disabilities in imaging speed and resolution, leveraging their abilities in pattern recognition and data analysis \cite{Hossain2024}.
DL techniques can enhance image quality for 0.55T MRI systems, boosting SNR and CNR to compete with higher field systems in image output. In higher field strengths DL can assist in handling extensive data, minimising artefacts, and enhancing image reconstruction processes for UHF MRI systems \cite{Hossain2024}, \cite{Arnold2023}, \cite{Vachha2021}.
Therefore, these advancements ensure better diagnostic accuracy, easier patient access, and increased efficiency in imaging, ultimately influencing the direction of MRI in clinical settings.

\section{Advances in Low-Field MRI - 0.55T}
\paragraph{}
In 2021, Siemens Healthineers achieved a significant milestone by receiving FDA clearance for its innovative 0.55T field strength MRI scanner, the High-V MRI \cite{mri_055}. This cutting-edge system combines advanced image processing techniques with computational methods to enhance its capabilities. 
The High-V scanner features a wide 80 cm bore aperture, purposefully designed to accommodate patients of larger bodies or those who experience claustrophobia, prioritising patient comfort and accessibility.
This system has a framework called Deep Resolve, which optimises the use of raw data acquired during the initial imaging process. Deep Resolve's algorithms construct individualised noise maps for each scan, identifying areas prone to noise corruption within the final image. 
This precise identification enables targeted measures to reduce noise, resulting in significant image quality improvements without the extension of scan times \cite{DL_mri_055}.
\paragraph{}
This section examines the applications of computational and advanced processing methods in the context of 0.55T MRI, highlighting recent progress and case studies in this field. 
The numerous articles discussed here have been categorised by body region to illustrate the diverse diagnostic potential of these enhanced low-field MRI approaches. 
Each sub-section presents the studies related to that category, which are also presented in table form, organised by author, referencing the article, objective, method and results of the study.
\paragraph{}
In 2023, Wang et al. \cite{Wang2023} developed a method for 2D turbo spin-echo (TSE) imaging, SPRING-RIO TSE, which compensates for concomitant gradient fields and B0 inhomogeneity at both 0.55T and 1.5T magnetic field strengths. 
To compensate for concomitant gradient fields and B0 inhomogeneity, they implemented gradient waveform modifications and reconstruction-based corrections. 
These modifications were designed to make up for the concomitant self-squared gradient terms and the phase increments induced by the B0 field throughout the reading.
Through simulations and tests using phantoms, the proposed modifications proved effective. 
Further validation with data from volunteer subjects confirmed that these modifications enabled high-quality images, enhanced SNR efficiency, and reduced the RF-specific absorption rate (SAR) compared to conventional turbo spin echo imaging techniques.
The study indicates that this new method, SPRING-RIO TSE, offers benefits, particularly in low-field MRI.

\subsection{Body Composition Study}
\paragraph{}
This subsection presents studies on the application of DL with MRI scans for body composition and abdominal studies. The studies explored here are shown in Table \ref{tab1}. 
Rusche et al. \cite{Rusche2022_confort} performed a comparative study in 2022 on subject comfort during MRI examination using the latest low-field MRI system with a wider bore aperture (0.55T) compared to the 1.5T MRI system.  
The research involved a group of 50 patients who had undergone an MRI scan on both machines involved in the survey. 
Those patients were required to assess their sense of space, noise level, and comfort as well as coil comfort and overall impression of the exam. The findings indicated that the majority of patients regarded, the 0.55T MRI scanner as better relating to sense of space, noise levels and general comfort. These noise levels were also measured and showed that 0.55T had markedly reduced noise for all body regions compared to 1.5T:\
\begin{itemize}
\item Brain MRI: 83.8 ± 3.6 dB (0.55T) vs. 89.3 ± 5.4 dB (1.5T), p = 0.04
\item Spine MRI: 83.7 ± 3.7 dB (0.55T) vs. 89.4 ± 2.6 dB (1.5T), p = 0.004
\item Hip MRI: 86.3 ± 5.0 dB (0.55T) vs. 89.1 ± 1.4 dB (1.5T), p = 0.04
\end{itemize}
\paragraph{}
The study found that the larger aperture and lower noise levels of 0.55T MRI scans contribute to patient's comfort during the procedure, reducing anxiety levels.
\paragraph{}
In 2023, Nayak et al. \cite{Nayak2023} investigated whether it would be possible to profile body composition using a whole-body MRI system operating at low-field strengths. 
Through scans on healthy adults, they refined parameters such as echo time, resolution, and inversion angle. A test-retest study with 10 volunteers confirmed the viability of the optimised protocol. 
Despite the differences in optimal parameters compared to higher field strengths, image quality remained comparable, without significant issues. Consequently, the study concludes that body composition MRI at 0.55T, with optimised parameters, is feasible for assessing conditions like metabolic diseases and obesity.
\paragraph{}
In 2024, Ramachandran et al. \cite{Ramachandran2024} made a comparative analysis of abdominal MRI images taken from healthy subjects and patients using a 0.55T MRI. 
The study involved 15 healthy subjects and 52 patients who were scanned by abdominal MRI scans at 0.55T. Additionally, images were taken of healthy individuals at 1.5T and the images were available for comparison on 28 of the 52 patients. 
Two radiologists independently assessed and scored the quality of the images using a four-level rating scale.
The images acquired at 0.55T were considered satisfactory, although some sequences may require further optimisation. The radiologists were able to respond to clinical queries for the majority of the patient cases. While the diagnostic quality was satisfactory, certain sequences at 0.55T received lower scores compared to those at higher field strengths. 
Moreover, the average scan time at 0.55T was found to be longer than that at higher field strengths, which may present a challenge in clinical settings.
\paragraph{}
Tian et al. \cite{Tian2024} introduced a real-time MRI technique at 0.55T using spiral out-in-out-in sampling combined with balanced steady-state free precession (bSSFP) to separate water and fat effectively. 
This method, tested on healthy volunteers, successfully visualised wrist ligament gaps during motion, enhanced the clarity of epicardial fat and coronary arteries in the heart, and improved the water-fat contrast for observing small bowel motility in the abdomen. 
The flexible sequence can be customised for various temporal and spatial resolutions, thereby demonstrating its adaptability for a multitude of clinical and research applications.
\paragraph{}
Dietzel et al. \cite{Dietzel2024} investigated the potential of using a low-field MRI system operating at 0.55T for breast imaging. 
The study involved a multiparametric breast MRI protocol applied to 12 participants, using a next-generation scanner (MAGNETOM Free.Max). The protocol included dynamic T1-weighted, T2-weighted, and diffusion-weighted sequences optimised for the low-field setting. 
Two experienced radiologists independently evaluated the images and confirmed that all examinations yielded diagnostic-quality images without significant artefacts. The typical imaging phenotypes were effectively visualised, and the total scan time ranged from 10 minutes and 30 seconds to 18 minutes and 40 seconds.
The results of the study indicated that low-field breast MRI at 0.55T is a viable option for clinical use, delivering diagnostic-quality images within a reasonable examination time. This approach is expected to result in reduced procedural costs and facilitate the integration of MRI systems directly within breast clinics, potentially enhancing accessibility and efficiency in breast cancer diagnostics.
\paragraph{}
Kim et al. \cite{Kim2024} proposed a novel approach to accelerate MR cholangiopancreatography (MRCP) scans involving DL-based reconstruction techniques at both 3T and 0.55T field strengths. 
The research team utilised retrospectively undersampled data acquired at 3T to train a variational neural network. They then evaluated the performance of their proposed method against conventional techniques like parallel imaging and compressed sensing.
Furthermore, they developed a self-supervised DL reconstruction method and tested their approach in both retrospective and prospective accelerated scenarios. 
The results demonstrated a reduction in the average acquisition time with the variational network while maintaining an image quality comparable to that of conventional acquisition methods. 
The variational network exhibited superior performance in preserving the sharpness and visibility of hepatobiliary ducts, with promising adaptability to MRCP at 0.55T. 
This study showed the potential of DL-based reconstruction to enhance the efficiency and quality of MRCP imaging, paving the way for accelerated clinical applications.

\begin{longtable}{|p{2cm}|p{4cm}|p{4.5cm}|p{4.5cm}|}
    \caption{Summary of Studies on Body Composition Low-Field MRI Systems.}
    \label{tab1} \\
    \hline
    \textbf{Author} & \textbf{Purpose} & \textbf{Methods} & \textbf{Results} \\
    \hline
    \endfirsthead

    \multicolumn{4}{c}
    {\tablename \thetable -- {Summary of Studies on Body Composition Low-Field MRI Systems.}} \\
    
    \hline
    \textbf{Author} & \textbf{Purpose} & \textbf{Methods} & \textbf{Results} \\
    \hline
    \endhead

    \multicolumn{4}{r}{{(Continues on next page)}} \\
    \endfoot

    \endlastfoot
        Rusche et al. (2022) \cite{Rusche2022_confort} & Compare patient comfort between 0.55T and 1.5T MRI systems. &   50 patients underwent scans on both systems;
        
          Two radiologists rated sense of space, noise, and comfort. &   0.55T rated better for space, comfort, and significantly lower noise levels. \\ \hline
        Nayak et al. (2023) \cite{Nayak2023} & Assess feasibility of 0.55T MRI for body composition profiling &   Scans on healthy adults; 
        
          Parameter optimisation; 
        
          Test-retest with 10 volunteers. &   0.55T MRI is feasible with optimised parameters, and achieved comparable image quality; 
        
          It is useful for metabolic diseases and obesity. \\ \hline
        Ramachandran et al. (2024) \cite{Ramachandran2024}& Compare abdominal MRI at 0.55T and 1.5T in healthy subjects and patients &   15 healthy subjects and 52 patients scanned on both MRI; 
        
          The scans were evaluated by two radiologists. &   0.55T images satisfactory with lower scores for some sequences, and longer scan time. \\ \hline
        Tian et al. (2024) \cite{Tian2024}& Introduce real-time MRI technique at 0.55T &   Tested on healthy volunteers using spiral sampling and bSSFP. &   Effective visualisation of wrist ligament gaps, epicardial fat, coronary arteries, and small bowel motility. \\ \hline
        Dietzel et al. (2024) \cite{Dietzel2024}& Investigate 0.55T MRI for breast imaging &   Multiparametric protocol on 12 participants; 
        
          The scans were evaluated by two radiologists. &   It has diagnostic-quality images, and there are no significant artefacts, which make it viable for clinical use. \\ \hline
        Kim et al. (2024) \cite{Kim2024} & Accelerate MRCP acquisitions using DL-based reconstruction techniques &   The variational network was trained using undersampled 3T data, self-supervised DL reconstruction, and both retrospective and prospective scenarios. &   The acquisition time was significantly reduced, while the image quality was maintained, and the visibility of hepatobiliary ducts was enhanced;
        
          The adaptability to 0.55T MRCP was superior. \\ \hline
\end{longtable}

\subsection{Musculoskeletal Study}
\paragraph{}
Low-field MRI has the capability to diagnose various musculoskeletal (MSK) conditions. Researchers have investigated the potential benefits and limitations of using low-field MRI systems for MSK imaging applications.
The studies explored here are presented in Table \ref{tab2}.
In 2023, Breit et al. \cite{Breit2023} assessed the potential of a 0.55T MRI system for lumbar spine imaging, comparing it to a standard 1.5T system. The study included 14 volunteers who underwent imaging on both systems, with additional sequences acquired at 0.55T using advanced post-processing techniques. Although the 0.55T images were rated slightly lower in quality compared to the 1.5T images for signal/contrast, resolution, and visualisation, the overall quality of the examination remains good, with a high level of agreement between radiologists. The application of advanced post-processing techniques at 0.55T data showed potential for enhancing image quality while also reducing the necessary acquisition time.
\paragraph{}
In 2024, Lavrova et al. \cite{Lavrova2024} conducted a study that compared the image quality, variability between reader's interpretations, and diagnostic effectiveness of standard lumbar spine MRI sequences acquired at 0.55T versus those acquired at the more conventional field strengths of 1.5T and 3T.  
The study involved 665 image series from 70 studies, which were evaluated by two neuroradiologists for overall imaging quality, the presence of artefacts, and the ability to visualise anatomical features using a four-point Likert scale. 
Despite some sequences at 0.55T receiving lower scores, all were considered acceptable for diagnostic use. There was a low level of agreement between readers on individual scores, but a high level of agreement on diagnoses, indicating the potential of 0.55T MRI for detecting routine spine pathologies.
\paragraph{}
The study by Schlicht et al. \cite{Schlicht2024} aimed to evaluate the performance of a DL-based post-processing technique for enhancing lumbar spine MRI scans acquired at 0.55T. 
The primary goals were to assess the resulting image quality and the potential reduction in acquisition time offered by this advanced method.
A comparison was made between standard DL-based reconstruction and advanced reconstruction (ADLR) in the context of lumbar spine imaging of 18 patients. 
A visual evaluation by radiologists using a Likert scale demonstrated superior image quality, resolution, and assessability of anatomical structures with the ADLR technique. A quantitative analysis revealed that ADLR resulted in higher contrast ratios in fluid-sensitive sequences. Furthermore, ADLR resulted in a significant reduction in image acquisition time, with a decrease of 44.4\% from 14:22 minutes to 07:59 minutes. 
These results show that advanced DL-based reconstruction algorithms can improve image quality and reduce the time needed to take a lumbar spine MRI at 0.55T.
\paragraph{}
Seifert et al. \cite{Seifert2024} evaluated the image quality and metal artefact intensity between 0.55T and 1.5T MRI in patients who had undergone posterior fusion surgery with spinal implants.
The study involved 50 patients and employed metal artefact reduction protocols used in clinical practice for the evaluation of images. 
The findings revealed that while contrast and resolution were rated lower at 0.55T compared to 1.5T MRI, the ability to evaluate the spinal canal was equally effective at both field strengths. However, mild susceptibility artefacts were observed at 0.55T. 
Furthermore, the study indicated that
while the visibility of anatomical structures near metal implants reduced as the implant length increased in 1.5T MRI scans, the visibility stayed consistent and unaffected by implant length in the 0.55T MRI scans.
This suggests that the latter may offer a superior assessment of near-metal anatomy, particularly in cases involving multi-level spinal implants.
\paragraph{}
In 2023, Chaudhari et al. \cite{Chaudhari2023} conducted a study investigating the feasibility of real-time MRI of the moving wrist using a high-performance 0.55T system. 
Conventional MRI systems are constrained by limitations in real-time imaging due to the generation of artefacts and power deposition. 
The study participants underwent wrist motion manoeuvres during MRI scans with temporal resolutions below 100 milliseconds. The objective of the study was to visualise tissues to evaluate dynamic wrist instability. The results demonstrated minimal distortion and image artefacts in real-time MRI scans, with static MRI providing a complement to the assessment. 
In conclusion, the study demonstrated that high-temporal-resolution real-time MRI in conjunction with static MRI is a viable approach with a 0.55T system, offering enhanced evaluation of wrist dysfunction and instability.
\paragraph{}
Lim et al. \cite{Lim2024} investigated the potential of real-time MRI to observe the dynamics of speech production using a 0.55T MRI system. 
The study involved healthy adult volunteers and employed a range of pulse sequences, including bSSFP and gradient-recalled echo. 
The results demonstrated that both imaging sequences were capable of effectively tracking and visualising the intricate tongue motion patterns during the rapid articulation of consonant-vowel syllables.
Nevertheless, bSSFP demonstrated superior SNR in the vocal tract articulators and offered enhanced image quality and SNR efficiency, particularly when utilising the shortest readout duration. This result indicates that bSSFP outperforms gradient-recalled echo in terms of image quality, even when compared to higher field strengths. 
The study highlights the potential of using high-performance 0.55T MRI systems for speech-production real-time MRI, with spiral bSSFP demonstrating the capacity to provide enhanced SNR efficiency and image quality, while avoiding the common artefacts.
\paragraph{}
Schmidt et al. \cite{Schmidt2023} conducted a study to assess the diagnostic image quality of knee MRI performed at a low magnetic field strength of 0.55T in comparison with the 1.5T MRI. They conducted MRI scans on 20 volunteers using both scanners. 
Two radiologists assessed the MRI sequences, scoring the overall image quality, level of image noise, and diagnostic utility using a five-point rating scale. Furthermore, the study involved assessing the potential presence of pathologies affecting the menisci, ligaments, and cartilage within the scanned regions.
The study demonstrated that while image quality for specific sequences at 0.55T was comparable to that of 1.5T MRI, it exhibited a slight decline for other sequences. 
Nevertheless, the diagnostic capability for meniscal and cartilage pathologies remained comparable between the two field strengths. Moreover, the tissue contrast ratios were comparable across both. The inter-observer agreement on image quality was satisfactory, particularly about the identification of pathologies. 
Therefore, the implementation of DL-reconstructed images at 0.55T produced knee MRI scans yielded image quality comparable to that obtained from the higher 1.5T field strength MRI scanners.
\paragraph{}
A complementary study by Donners et al. \cite{Donners2023} was conducted to assess the efficacy of the new-generation 0.55T knee MRI in comparison to 3T knee MRI for patients experiencing acute trauma and knee pain. 
The research involved 25 patients presenting with symptoms to obtain knee MRIs on both 0.55T and 3T systems. DL image reconstruction algorithm was employed only on the 0.55T system, resulting in enhanced image quality. 
Two radiologists independently reviewed all images, assessing image quality parameters, detecting MRI findings, and evaluating confidence levels in reporting these findings. The study demonstrated that while image quality was rated higher for 3T MRI, there was agreement between 0.55T and 3T MRI in detecting various knee lesions. 
Nevertheless, there was a lower degree of agreement and reader confidence in the detection of low-grade cartilage and meniscal lesions in the 0.55T MRI compared to the 3T MRI. 
This suggests that while the 0.55T knee MRI system, combined with DL image reconstruction, can reliably detect and grade joint lesions in symptomatic patients, it may face challenges in accurately evaluating low-grade cartilage and meniscal lesions when compared to the higher 3T MRI field strength.
\paragraph{}
Bhattacharjee et al. \cite{Bhattacharjee2023} addressed the feasibility of applying DL segmentation models, originally trained on 3T MRI data, directly to low-field 0.55T knee MRI scans of healthy controls. The objective of this study is to evaluate the performance of DL algorithms for quantifying knee biomarkers at 0.55T compared to 3T. Seven healthy subjects underwent bilateral knee MRI acquisitions on both 3T and 0.55T scanners. 
DL segmentation models were applied to segment knee bones and cartilage, and the segmentations were qualitatively and quantitatively evaluated. 
The results showed that the DL models exhibited a satisfactory capability to accurately segment and delineate bones and cartilaginous structures at 0.55T, although segmentation quality varied between different knee regions. 
While bone segmentation performance was generally better at 3T, cartilage segmentation was comparable between 0.55T and 3T, particularly for the femoral and tibial regions. 
However, the patellar region posed challenges, likely due to lower SNR at 0.55T. Despite these challenges, the study demonstrated the technical feasibility of using DL segmentation models trained on higher-field MRI data for knee imaging at 0.55T. This approach could facilitate knee biomarker quantification in low-field MRI settings.
\paragraph{}
In 2024, Plesniar et al. \cite{Plesniar2024} assessed 
the diagnostic accuracy of 0.55T MRI in identifying periprosthetic joint infection (PJI) among patients with symptomatic total hip arthroplasty (THA). The analysis included MRI scans from three groups of patients: Group A consisted of individuals with confirmed PJI, Group B included patients with non-infected THA, and Group C was a subgroup of Group B with aseptic loosening of the implant. 
The radiologists evaluated the MRI scans independently, identifying various markers of infection. Capsule oedema, intramuscular oedema, and joint effusion were identified as the most effective discriminators for the diagnosis of PJI, exhibiting high sensitivity and specificity when tested individually and in parallel. 
Moreover, these markers demonstrated significance in differentiating between PJI and aseptic loosening. The results indicate that 0.55T MRI, particularly with its capacity to identify oedema in the joint capsule and adjacent muscles, has the potential to to aid in detecting PJI in symptomatic patients who have undergone THA.
\paragraph{}
Wiesmueller et al. \cite{Wiesmueller2023} aimed to evaluate the usefulness of 0.55T MRI for diagnosing vestibular schwannomas (VS) in the internal auditory canal (IAC). The IAC of 56 patients with unilateral VS was imaged using both 1.5T and 0.55T MRIs. 
Several MRI sequences were independently assessed by two radiologists for image quality, visibility of VS, diagnostic confidence and artefacts that may be present while a second independent reading involved a direct comparison of images at both field strengths. 
The findings revealed that there was no substantial difference in image quality between the scans when considering factors such as visibility, diagnostic confidence, or the presence of artifacts. Additionally, direct comparison of the images did not indicate any significant difference in the ability to discern lesions or the level of confidence in making diagnoses.
Consequently, low-field MRI demonstrated satisfactory diagnostic image quality and proved to be feasible in evaluating VS within IAC.
\paragraph{}
According to Kopp et al. \cite{Kopp2024}, the 0.55T MRI for temporomandibular disorders (TMDs) was evaluated in comparison with the 1.5T standard MRI system. 
Temporomandibular Joint Disorder is a condition that is common causing functional restrictions and pains, therefore it often needs an MRI examination to assess the temporomandibular joint (TMJ) in detail. 
Low-field MRI has not yet been tested as a means of evaluating TMJ despite being a commonly used method in TMJ assessment using MRI imaging at 1.5T and 3T. On the same day, 17 patients with suspected intra-articular TMDs underwent both 0.55T and 1.5T MRI scans.
Two senior readers independently assessed the image quality of the MRI scans while focusing on disc morphology and position as well as osseous joint morphology for each TMJ: by way of a four-point Likert scale, image quality and the degree of artefacts were evaluated for both field strengths.
The results indicate that compared to those using 1.5T MRI, image quality at 0.55T was poor for disc morphology, and osseous joint morphology but almost similar for disc position between two field strengths. 
However, most cases were diagnosed correctly at both 0.55T and 1.5T since minor image artifacts were more common at 0.55T in comparison with 1.5T. This suggests that 0.55T MRI may help in diagnosing TMDs.

\begin{longtable}{|p{1.5cm}|p{4.5cm}|p{4.5cm}|p{4.5cm}|}
    \caption{Summary of MSK Studies on Low-Field MRI Systems.}
    \label{tab2}\\
    \hline
    \textbf{Author} & \textbf{Purpose} & \textbf{Methods} & \textbf{Results} \\
    \hline
    \endfirsthead

    \multicolumn{4}{c}%
    {\tablename\ \thetable\ -- {Summary of MSK Studies on Low-Field MRI Systems.}} \\
    \hline
    \textbf{Author} & \textbf{Purpose} & \textbf{Methods} & \textbf{Results} \\
    \hline
    \endhead

    \multicolumn{4}{r}{{(Continues on next page)}} \\
    \endfoot

    \hline
    \endlastfoot
        Breit et al. (2023) \cite{Breit2023} & Evaluate the potential of 0.55T MRI for imaging the lumbar spine &   Compared 0.55T to 1.5T with 14 volunteers using advanced post-processing. &   0.55T classified slightly lower but overall good quality, with potential for enhanced image quality with reduced acquisition time. \\ \hline
        Lavrova et al. (2024) \cite{Lavrova2024} & Compare the quality and performance of low-field and high-field lumbar spine MRI scans &   665 image series from 70 studies assessed by two neuroradiologists. &   0.55T sequences considered acceptable, high agreement on diagnoses despite lower scores. \\ \hline
        Schlicht et al. (2024) \cite{Schlicht2024} & Assess the effectiveness of DL-based post-processing in lumbar spine MRI at 0.55T &   Compared conventional with ADLR on 18 patients. &   ADLR improved image quality, and resolution, and reduced acquisition time by 44.4\%. \\ \hline
        Seifert et al. (2024) \cite{Seifert2024} & Compare image quality and metal artefact severity between 0.55T and 1.5T MRI in patients with spinal implants &   50 patients with metal artefact reduction protocols. &   0.55T had lower contrast/resolution but equal spinal canal assessment, and better near-metal anatomy visibility. \\ \hline
        Chaudhari et al. (2023) \cite{Chaudhari2023} & Investigate real-time MRI of the moving wrist using 0.55T &   Wrist motion manoeuvres with temporal resolutions below 100 ms. &   Minimal distortion and artefacts, suitable for improved assessment of wrist dysfunction. \\ \hline
        Lim et al. (2024) \cite{Lim2024} & Investigate real-time MRI for speech production dynamics using 0.55T &   Healthy volunteers, using bSSFP and gradient-recalled echo sequences. &   bSSFP showed superior SNR and image quality, effective for capturing tongue movements. \\ \hline
        Schmidt et al. (2023) \cite{Schmidt2023} & Evaluate diagnostic image quality of knee MRI at 0.55T vs. 1.5T &   MRI on 20 volunteers, assessed by two radiologists. &   Comparable diagnostic capability for meniscal/cartilage pathologies, a slight decline in specific sequences. \\ \hline
        Donners et al. (2023) \cite{Donners2023} & Assess new-generation 0.55T knee MRI vs. 3T in acute trauma/knee pain patients &   25 patients;
        
          DL image reconstruction at 0.55T. &   Agreement in detecting various lesions, lower agreement/confidence in low-grade cartilage/meniscal lesions. \\ \hline
        Bhattacharjee et al. (2023) \cite{Bhattacharjee2023} & Evaluate DL segmentation models trained on 3T MRI data on 0.55T knee MRI scans &   Develop a DL segmentation method for knee bones and cartilages;
        
          7 healthy subjects underwent bilateral knee MRI on 3T and 0.55T scanners.  &   Comparable cartilage segmentation could be achieved between 0.55T and 3T. 
        
          Reasonable bone segmentation was possible at 0.55T, and there was potential for the quantification of biomarkers in the knee in low-field MRI. \\ \hline
        Plesniar et al. (2024) \cite{Plesniar2024} & Assess the accuracy of 0.55T MRI in diagnosing PJI in THA patients &   MRI scans from patients with PJI, noninfected THA, and aseptic loosening. &   High sensitivity/specificity for identifying PJI, effective in differentiating from aseptic loosening. \\ \hline
        Wiesmueller et al. (2023) \cite{Wiesmueller2023} & Assess 0.55T MRI for diagnosing vestibular schwannomas in IAC &   MRI on 56 patients with unilateral VS, evaluated by two radiologists. &   Comparable image quality and diagnostic confidence to 1.5T, feasible for VS evaluation. \\ \hline
        Kopp et al. (2024) \cite{Kopp2024} & Evaluate 0.55T MRI for TMD assessment vs. 1.5T &   MRI on 17 patients with suspected TMDs, evaluated by two readers. &   Lower image quality at 0.55T, but sufficient diagnostic confidence in the majority of cases. \\ \hline
\end{longtable}

\subsection{NeuroImaging Study}
\paragraph{}
This subsection presents studies on the application of DL with brain MRI scans for neuroimaging purposes. The studies explored here are presented in Table \ref{tab3}.
In 2021, Campbell-Washburn et al. \cite{Campbell-Washburn_MRF} investigated the feasibility of using magnetic resonance fingerprinting (MRF) on a high performance 0.55T MRI system for rapid multi-parametric quantification. 
MRF is known for its versatility and ability to provide quantitative imaging at multiple field strengths. 
The study aims to demonstrate the potential of MRF for whole brain mapping on a modern 0.55T MRI system, which could improve examination efficiency and enable multi-contrast imaging on lower field strength systems. 
A whole-brain 3D stack-of-spirals fast imaging with steady-state precession (FISP) MRF sequence was implemented and quantification was validated using a standardised MRI phantom. 
The results showed strong correlations between MRF values and standard phantom measurements, with high precision despite the lower SNR at 0.55T. 
The study concludes that the proposed method is feasible at 0.55T, suggesting the potential for high-quality imaging at lower cost, making it attractive for various imaging scenarios.
\paragraph{}
In 2024, Kung et al. \cite{Kung2024} conducted a study to assess the feasibility of performing diffusion tensor brain imaging at a lower magnetic field strength of 0.55T compared to the more commonly used 3T. 
Data were collected from a small group of five healthy subjects using both 0.55T and 3T scanners, with a focus on achieving high-resolution imaging with adequate SNR. To overcome the challenge of reduced SNR at 0.55T, a signal enhancement technique called SNR-enhancing joint reconstruction was applied. 
The study found that after applying this reconstruction method, the diffusion tensor parameters obtained at 0.55T correlated strongly with those obtained at 3T, demonstrating the feasibility of performing high-resolution diffusion MRI at lower field strengths. 
In addition, test-retest analysis indicated improved repeatability of diffusion tensor parameters with the SNR-enhancing reconstruction at 0.55T. 
These results suggest that, with appropriate noise reduction strategies, high-quality diffusion MRI of the human brain is achievable at 0.55T.

\begin{longtable}{|p{1.5cm}|p{4.5cm}|p{4.5cm}|p{4.5cm}|}
    \caption{Summary of Studies on NeuroImaging on Low-Field MRI Systems.}
    \label{tab3}\\
    \hline
    \textbf{Author} & \textbf{Purpose} & \textbf{Methods} & \textbf{Results} \\
    \hline
    \endfirsthead

    \multicolumn{4}{c}%
    {\tablename\ \thetable\ -- {Summary of Studies on NeuroImaging on Low-Field MRI Systems.}} \\
    \hline
    \textbf{Author} & \textbf{Purpose} & \textbf{Methods} & \textbf{Results} \\
    \hline
    \endhead

    \multicolumn{4}{r}{{(Continues on next page)}} \\
    \endfoot

    \hline
    \endlastfoot
        Campbell-Washburn et al. (2021) \cite{Campbell-Washburn_MRF} & Investigate the feasibility of MRF for rapid multi-parametric quantification on 0.55T MRI &   Implemented 3D stack-of-spirals FISP MRF sequence, and validated with standardised MRI phantom. &   Strong correlation with standard measurements and high precision;
        
          Despite lower SNR, it shows potential for high-quality imaging at a lower cost. \\ \hline
        Kung et al. (2024) \cite{Kung2024} & Investigate the feasibility of diffusion tensor brain imaging at 0.55T vs. 3T &   Data from five healthy subjects, and applied SNR-enhancing joint reconstruction &   Strong correlation of diffusion tensor parameters between 0.55T and 3T, improved repeatability with SNR-enhancing reconstruction. \\ \hline
\end{longtable}

\subsection{CardioVascular Study}
\paragraph{}
This subsection presents studies on the application of DL with MRI scans of the heart and blood vessels. The studies explored here are shown in Table \ref{tab4}.
In 2020, Restivo et al. \cite{Restivo2020} investigated the benefits of using a 0.55T MRI for cardiac cine imaging. 
The researchers aimed to improve the SNR efficiency of long-read acquisitions, focusing on spiral and echoplanar imaging (EPI) techniques. They developed two high-efficiency bSSFP acquisition methods, a spiral in-out acquisition and an EPI acquisition incorporating cancellation techniques. The study included simulations, phantom images and images of 12 healthy volunteers to evaluate SNR and image quality. 
The results demonstrated that the spiral in-out bSSFP method performed well and produced high-quality images, while EPI bSSFP was more sensitive to movement and flow artefacts. The proposed method at 0.55T achieved a significant increase in myocardial SNR compared to the reference acquisition. 
The study demonstrated that the efficient spiral in-out bSSFP technique provides high-quality cardiac cine images and significant SNR recovery using a high-performance 0.55T MRI system.
\paragraph{}
Varghese et al. \cite{Varghese2023} developed a comprehensive cardiovascular magnetic resonance (CMR) exam protocol on a 0.55T MRI system, to overcome the challenges presented by lower field strength and slower gradient performance. 
The researchers constructed a prototype CMR exam on an ultra-wide bore 0.55T MRI system by leveraging compressed sensing-based techniques to compensate for the inherently lower SNR and slower imaging times. 
The study demonstrated that compressed sensing-based techniques enabled the acquisition of images of sufficient quality for the assessment of cardiovascular structure, function, and flow.
These techniques also enabled spatial and temporal resolution suggestions comparable to those achievable with higher field strength systems.
\paragraph{}
In 2024, McGrath et al. \cite{McGrath2024} investigated the use of bSSFP in a 0.55T low-field MRI system for cardiac assessment. 
The researchers employed a free-running radial acquisition technique in conjunction with cardiac and respiratory auto-regulation methods to obtain reliable blood flow measurements and generate high-quality images of the heart. By implementing a phase-contrast bSSFP sequence in 0.55T and 1.5T MRI systems, the researchers were able to exploit the flow sensitivity of bSSFP and the phase cancellation properties. 
This approach enabled tests to be conducted without the need for ECG, thereby reducing examination times while maintaining accuracy. 
The \textit{in vivo} data collected from healthy subjects demonstrated that the bSSFP method with phase contrast is comparable to measurements obtained with gradient-echo techniques, thereby confirming its viability in low-field MRI systems.
\paragraph{}
The ability of 12 lead Electrocardiogram (ECG) monitoring in a 0.55T MRI system for improved monitoring of the heart during an MRI scan was evaluated by Kolandaivelu et al. \cite{Kolandaivelu2024}. The 12-lead ECG is a basic instrument for observing cardiac ischemia and arrhythmia in many hospital settings. 
However, the strong magnetic field generated by the MRI equipment can distort the ECG signal and make it less understandable due to its non-interpretability.
The objective of this study was to assess whether a 0.55T MRI scanner could provide superior ECG interpretation compared to higher field strengths. 
Experiments were conducted on pigs and healthy human volunteers, with ECGs being recorded inside different MRI scanners at various positions relative to the scanner. The study was designed to assess the occurrence and extent of ECG error and variation, with a particular focus on clinically relevant features. 
The results demonstrated that ECG distortion was significantly lower at 0.55T compared to higher field scanners, with distortion approaching acceptable levels even at the scanner table's home position. 
Consequently, the 0.55 T MRI scanner can facilitate ECG interpretation and reduce the need for uncomfortable patient repositioning, thus increasing the feasibility of ECG monitoring during interventional MRI procedures.
\paragraph{}
Subsequent studies have concentrated on interventional cardiac MRI procedures. Basar et al. \cite{Basar2021} conducted a study to analyse the susceptibility artefacts created by metallic markers and cardiac catheterisation devices in a 0.55T MRI system. These artefacts play a pivotal role in the visualisation of passive devices during MR-guided catheterisation. 
The researchers examined the field strength and orientation dependence of these artefacts to gain insight into their variations with different MRI field strengths.
Gradient echo images were employed at 0.55T, 1.5T, and 3T to quantify the volume of artefacts produced by various types of passive susceptibility markers made from nitinol and stainless steel. Furthermore, the visibility of catheterisation devices currently available for clinical use at 0.55T and 1.5T was compared in phantoms and \textit{in vivo} using real-time bSSFP imaging.
The results demonstrated that the size of susceptibility artefacts is dependent on field strength and orientation. It is noteworthy that stainless steel markers with high susceptibility produced field-independent artefacts at various field strengths, suggesting magnetic saturation below 0.55T.
Consequently, the size of the artefacts can be adjusted to ensure consistent imaging signatures of interventional devices at different MRI field strengths.
\paragraph{}
Rusche et al. \cite{Rusche2022_stroke} conducted a comparative study to assess the effectiveness of 0.55T MRI in imaging stroke. 
The study involved 27 participants, including stroke patients and control subjects, who underwent brain scans with both types of scanners. The data was analysed in three stages: qualitative assessment, lesion detection, and segmentation of lesion volume. 
The results demonstrated that the detection of lesions in diffusion-weighted imaging (DWI) and apparent diffusion coefficient (ADC) sequences was equivalent between the two scanners. The detection of lesions was also comparable. Noise levels were lower in the 0.55T sequences for DWI/ADC, but the 1.5T sequences were superior in other aspects. There was no difference in infarct volume between the two scanners. 
Thereby, low-field MRI at 0.55T could be a viable alternative for stroke imaging.
\paragraph{}
In a recent publication, Yildirim et al. \cite{Yildirim2023} presented a novel interventional MRI guidewire designed for cardiovascular catheterisation at 0.55T. The guidewire is constructed from a metallic material and exhibits a continuous image profile from its shaft to its tip. 
The researchers conducted electromagnetic simulations and real-time MRI visibility tests to ensure that the guidewire's shaft and tip exhibited continuous visibility. Subsequent tests of the guidewire's RF-induced heating and mechanical performance demonstrated its safety and effectiveness. The detachable connector facilitates the replacement of the catheter while maintaining the position of the guidewire, thereby enhancing its utility in clinical settings. 
This innovative design allows for the safe navigation of luminal structures and cardiac chambers, offering similar performance to X-ray guidewires during MRI catheterisation. 
\paragraph{}
In 2024, Uzun et al. \cite{Uzun2024} introduced a new method for tracking MRI intervention devices using markers printed with conductive ink, which are controlled by alternating current. 
The efficacy of the markers was evaluated on MRI-compatible specimens that are usually difficult to visualise during MRI scans. The visibility of the markers was evaluated using gradient echo, bSSFP and turbo spin-echo sequences with variable current parameters (amplitude, frequency) on 0.55T and 1.5T MRI scanners. A current supply circuit compatible with MRI technology was developed to regulate the current applied to the markers. 
The results demonstrated that adjusting the current parameters enhanced the visibility of the markers in all three sequences, thereby increasing their visibility. Furthermore, the study demonstrated that this method effectively increased the visibility of a customised nitinol needle in both \textit{in vivo} and post-mortem animal experiments. 
In conclusion, current-controlled markers printed in conductive ink represent a simple and effective solution for tracking MRI intervention devices, adaptable to various parameters of the pulse sequence by adjusting the applied current.

\begin{longtable}{|p{1.5cm}|p{4.5cm}|p{4.5cm}|p{4.5cm}|}
    \caption{Summary of Cardiovascular Studies on Low-Field MRI Systems.}
    \label{tab4}\\
    \hline
    \textbf{Author} & \textbf{Purpose} & \textbf{Methods} & \textbf{Results} \\
    \hline
    \endfirsthead

    \multicolumn{4}{c}%
    {\tablename\ \thetable\ -- {Summary of Cardiovascular Studies on Low-Field MRI Systems.}} \\
    \hline
    \textbf{Author} & \textbf{Purpose} & \textbf{Methods} & \textbf{Results} \\
    \hline
    \endhead

    \multicolumn{4}{r}{{(Continues on next page)}} \\
    \endfoot

    \hline
    \endlastfoot
        Restivo et al. (2020) \cite{Restivo2020} & Investigate benefits of 0.55T MRI with high-performance hardware for cardiac cine imaging &   Developed spiral in-out and EPI bSSFP acquisition methods, and it was evaluated in simulations with phantom images, and images of 12 healthy volunteers. &   Spiral in-out bSSFP method produced high-quality images with a significant increase in myocardial SNR. \\ \hline
        Varghese et al. (2023) \cite{Varghese2023} & Develop comprehensive CMR exam protocol on 0.55T MRI &   Prototype CMR on ultra-wide bore 0.55T MR system using compressed sensing-based techniques. &   Enabled acquisition of sufficient quality images for cardiovascular assessment, with comparable spatial and temporal resolution to higher field strengths. \\ \hline
        McGrath et al. (2024) \cite{McGrath2024} & Investigate the use of bSSFP in 0.55T MRI for cardiac assessment &   Free-running radial acquisition technique with cardiac and respiratory autoregulation methods. &   It was comparable to gradient-echo techniques, presenting reduced examination times, reliable blood flow measurements, and high-quality heart images. \\ \hline
        Kolandaivelu et al. (2024) \cite{Kolandaivelu2024} & Evaluate the performance of 12-lead ECG monitoring in 0.55T MRI scanner for improved cardiac monitoring during MRI &   Experiments on pigs and healthy volunteers;
        
          Assessment of ECG error and variation inside different MRI scanners. &   It showed significantly lower ECG distortion at 0.55T, facilitated interpretation, and reduced the necessity for patient repositioning. \\ \hline
        Basar et al. (2021) \cite{Basar2021} & Analyse sensitivity artefacts of metallic markers and cardiac catheterisation devices in 0.55T MRI &   Quantification of artefact volume from passive susceptibility markers was performed on gradient echo images at 0.55T, 1.5T, and 3T. &   Artefact size dependent on field strength and orientation;
        
          Stainless steel markers produced field-independent artefacts and showed consistent imaging signatures at different field strengths. \\ \hline
        Rusche et al. (2022) \cite{Rusche2022_stroke} & Assess effectiveness of 0.55T MRI in imaging stroke &   Brain scans of 27 participants;
        
          Qualitative assessment, lesion detection, and segmentation of lesion volume of the scans. &   Detection of lesions in DWI and ADC sequences was equivalent to 1.5T;
        
          At 0.55T it had lower noise levels and no significant difference in infarct volume. \\ \hline
        Yildirim et al. (2023) \cite{Yildirim2023} & Present novel interventional MRI guidewire for cardiovascular catheterisation at 0.55T &   Designed a guidewire, tested for visibility, RF-induced heating, and mechanical performance. &   Continuous visibility of guidewire shaft and tip was guaranteed;
        
          It also showed safe navigation through luminal structures. \\ \hline
        Uzun et al. (2024) \cite{Uzun2024} & Introduce method for tracking MRI intervention devices using conductive ink-printed markers &   The efficacy of the markers was evaluated on MRI-compatible samples using varying pulse sequence parameters on 0.55T and 1.5T MRI. &   The enhanced marker visibility, which is enabled by adjustable current parameters, allows for the increased visibility of a custom-designed nitinol needle in both \textit{in vivo} and post-mortem experiments. \\ \hline
\end{longtable}

\subsection{Pulmonary Study}
\paragraph{}
This subsection explores studies on the application of DL in MRI examinations of the lungs and airways. The studies explored here are shown in Table \ref{tab5}.
In 2020, Campbell-Washburn et al. \cite{Campbell} examined the potential of a 0.55T MRI for imaging lung diseases and compared its image quality with that of clinical CT scans. Despite the limited use of MRI in lung imaging compared to CT, recent advancements in ultrashort echo time (UTE) gradient-echo imaging show promise for generating high-quality T1-weighted lung images.
The study describes a 0.55T low-field MRI system with extended upper-field homogeneity, which has the potential to reduce susceptibility artefacts in lung imaging. The study involved 24 participants with common lung anomalies. 
The researchers obtained high-quality structural MRI images of the lung in an average time of 11 minutes. 
The results demonstrated the ability to effectively detect a range of lung abnormalities in a low-field MRI. However, it proved more challenging to distinguish between diffuse diseases. 
Nevertheless, the study demonstrates that high-performance 0.55T MRI is promising for assessing common lung diseases, with strong correlations observed between nodule sizes measured on CT and MRI.
\paragraph{}
Bhattacharya et al. \cite{Bhattacharya2020} proposed an ML method for segmenting cystic structures in the lung using MRI, with a focus on patients with lymphangioleiomyomatosis (LAM). Taking advantage of advances in low-field MRI technology, the study offers a radiation-free alternative to conventional CT imaging. 
A DL pipeline incorporating generative adversarial networks (GANs) and modified UNet architectures is used to accurately identify lung cysts and tissues in MRI scans. By using a CycleGAN for MRI to CT synthesis and subsequent segmentation with a modified residual UNet, the approach achieved good results, validated by high average Dice scores. 
The study included data from 65 LAM patients, with MRI and CT scans used for training and validation, demonstrating the impact of MRI-based cyst quantification methods. 
These findings have the potential to transform diagnostic practice by providing safer and more efficient tools for cyst analysis in clinical settings.
\paragraph{}
In 2022, a study was conducted by Azour et al. \cite{Azour2022} to assess the efficacy of non-contrast 0.55T MRI in detecting and characterising ground-glass opacities and fibrosis in post-COVID patients. 
The MRI images were compared to clinically acquired chest CT images.
The study involved 64 participants with a history of COVID-19 pneumonia, with 20 of them also undergoing chest CT scans. Two thoracic radiologists reviewed and scored the images for various parameters, including the presence, extent, distribution, and presence of fibrosis-like features. 
The inter-reader agreement for the CT images was found to be high, while that for the MRI images was found to be reasonable to moderate. The degree of agreement between CT and MRI scans was found to range from moderate to substantial. Despite the challenges associated with lung MRI, such as signal characteristics and respiratory motion, low-field 0.55T MRI has demonstrated potential for the detection and characterisation of lung abnormalities.
\paragraph{}
Javed et al. \cite{Javed2022} developed a high-resolution imaging technique for pulmonary imaging at 0.55T. Their approach involved using a stack-of-spirals UTE sequence for continuous data acquisition. 
The acquired data was grouped into a stable respiratory phase using self-navigator signals. 
The corrections for trajectory errors and concomitant field artefacts were applied during image reconstruction. The study evaluated image quality using SNR, image sharpness, and qualitative reader scores in healthy volunteers, patients with COVID-19 infection, and patients with lung nodules. 
The technique yielded diagnostic-quality images with good SNR. The technique incorporates respiratory binning, which enhances image sharpness, and field corrections, which facilitate the observation of anatomical details in areas that are not situated in the centre of the image.
The technique maintained image quality despite a slight reduction in SNR for shorter scan times. Inline image reconstruction and artefact correction can be achieved in less than five minutes. 
Consequently, the technique developed integrates several advanced methods to enable the acquisition of high-quality pulmonary images with a 1.75 mm isotropic resolution on a 0.55T MRI system within a scan time of 8.5 minutes. It employs an efficient stack-of-spirals imaging approach, coupled with respiratory binning to account for breathing motion, field correction to mitigate artifacts arising from field inhomogeneities, and trajectory correction to compensate for errors in the data acquisition trajectory.
\paragraph{}
In 2023, Seemann et al. \cite{Seemann2023} introduced a time-resolved 3D MRI technique capable of measuring the transient dynamics of water in the lungs during both resting and physically performing conditions.
The objective was to detect early signs of heart failure manifested through exercise-induced dyspnea due to lung water accumulation. 
The study was conducted at 0.55T and involved 15 healthy subjects and two patients with heart failure. 
The results demonstrated that lung water density (LWD) exhibited an increase during exercise, particularly in healthy individuals, while patients exhibited slower accumulation rates. Furthermore, regional differences in the distribution of LWD between the anterior and posterior lungs were observed. 
These findings provide insights into the variations in lung water dynamics under different physiological conditions.
\paragraph{}
In their study, Li et al. \cite{Li2023} developed an improved low-field lung MRI technique using 0.55T MRI that utilises the transverse relaxation rates (R2 and R'2) of the lung parenchyma. 
The study was conducted on healthy adult volunteers and employed a bespoke single-slice spin-echo multi-echo (ES-MCSE) pulse sequence triggered by ECG on a prototype 0.55T whole-body MRI scanner. 
The results demonstrated that while the average R2 was comparable to that of 1.5T MRI, the R'2 was notably lower, indicating the potential for accurate measurements in 0.55T MRI. 
This research highlights the importance of the design and optimisation of new imaging methodologies for low-field lung MRI.

\begin{longtable}{|p{1.5cm}|p{4.5cm}|p{4.5cm}|p{4.5cm}|}
    \caption{Summary of Pulmonary Studies on Low-Field MRI Systems.}
    \label{tab5}\\
    \hline
    \textbf{Author} & \textbf{Purpose} & \textbf{Methods} & \textbf{Results} \\
    \hline
    \endfirsthead

    \multicolumn{4}{c}%
    {\tablename\ \thetable\ -- {Summary of Pulmonary Studies on Low-Field MRI Systems.}} \\
    \hline
    \textbf{Author} & \textbf{Purpose} & \textbf{Methods} & \textbf{Results} \\
    \hline
    \endhead

    \multicolumn{4}{r}{{(Continues on next page)}} \\
    \endfoot

    \hline
    \endlastfoot
        Campbell-Washburn et al. (2020) \cite{Campbell} & Investigate potential of 0.55T MRI for imaging lung diseases &   0.55T MRI system with extended upper field homogeneity;
        
          The study included 24 participants with lung anomalies. &   The 0.55T MRI demonstrated high-quality structural images and effective detection of lung abnormalities, but not to distinguish between diffuse diseases;
        
         There was a strong correlation between the MRI and CT nodule sizes. \\ \hline
        Bhattacharya et al. (2020) \cite{Bhattacharya2020} & Segment lung cystic structures in LAM patients using MRI &   Development of a DL pipeline with GANs and modified UNet architectures using MRI and CT data from 65 LAM patients. &   It demonstrated high average Dice scores, potential for radiation-free cyst quantification, clinical relevance for safer and more efficient diagnostic tools. 
        \\ \hline
        Azour et al. (2022) \cite{Azour2022} & Assess the efficacy of non-contrast 0.55T MRI in detecting and characterising post-COVID lung abnormalities &   MRI scans of 64 individuals with a history of COVID-19 pneumonia were compared with chest CT images. &   Moderate to substantial agreement between CT and MRI images, showing potential for detection and characterisation of lung abnormalities. \\ \hline
        Javed et al. (2022) \cite{Javed2022} & Develop high-resolution imaging sequence for pulmonary imaging at 0.55T &   The efficacy of the Stack-of-spirals UTE sequence, respiratory binning, field and trajectory corrections was evaluated in healthy volunteers, patients with COVID-19, and patients with lung nodules. &   The images produced are of diagnostic quality, with a good SNR and maintained image quality despite shorter scan times;
        
          The images have a high-quality with an isotropic resolution of 1.75 mm and can be produced in 8.5 minutes. \\ \hline
        Seemann et al. (2023) \cite{Seemann2023} & Measure transient lung water dynamics during rest and exercise stress &   Developed a time-resolved 3D MRI technique;
        
          The study involved 15 healthy subjects and two heart failure patients. &   A notable increase in LWD has been observed during exercise, with a slower accumulation observed in heart failure patients;
        
          Regional differences in lung water distribution have been identified. \\ \hline
        Li et al. (2023) \cite{Li2023} & Develop improved low-field lung MRI technique using 0.55T MRI &   A single-slice spin-echo multi-echo pulse sequence was conducted on healthy adult volunteers, triggered by an ECG. &   The average R2 value is comparable to that of a 1.5T MRI, while the lower R2 value indicates the potential for accurate measurements. \\ \hline
\end{longtable}

\subsection{Fetus Study}
\paragraph{}
This subsection explores studies on the application of DL in fetal and pregnancy MRI scans. The studies explored here are presented in Table \ref{tab6}.
In 2023, Ponrartana et al. \cite{Ponrartana2023} investigated fetal assessment with the 0.55T MRI system. 
The researchers presented an approach for achieving diagnostic-quality fetal MRI. In the past, fetal MRI was conducted on higher-field systems, such as 1.5T or 3T. Nevertheless, the advent of low-field strength 0.55T MRI systems presents several advantages, including enhanced field homogeneity, and a decrease in acoustic noise. These advantages enhance patient comfort and safety. The study outlines the experimental methods employed, including data collection on a 0.55T MRI system using fast sequence acquisition to mitigate the impact of fetal movement.
The fetal MRI protocol incorporates a range of sequences, in addition to supplementary techniques such as DWI, hydrography, and cine imaging. Collectively, these techniques enable a comprehensive assessment of fetal health. 
The research demonstrates the potential of low-field MRI systems for conducting high-quality fetal MRI, providing a promising avenue for fetal imaging with an improved patient experience.
\paragraph{}
Verdera et al. \cite{Verdera2023} assessed the reliability and feasibility of low-field fetal MRI to evaluate anatomical and functional measurements in pregnant women.
The study employed a 0.55T MRI scanner and a 20-minute protocol. The study included healthy pregnant women and women with pregnancy-related abnormalities, with a total of 79 fetal MRI scans performed. 
The 20-minute protocol, which included fast-spin-echo anatomical sequences, quantitative T2* and diffusion, produced key measurements such as biparietal diameter, transcerebellar diameter, lung volume and cervical length. 
These measurements were assessed by two radiologists and an obstetrician experienced in MRI.
The results demonstrated concordance with large cross-sectional control studies for the anatomical measurements, and high inter-observer correlations were observed for the biparietal diameter. The functional characteristics, including the placental and brain T2* values and the placental ADC values, demonstrated a robust correlation with gestational age. 
The study demonstrated that the 20-minute low-field fetal MRI protocol reliably produces structural and functional measurements of the fetus and placenta during pregnancy.
\paragraph{}
In addition, Slator et al. \cite{Slator2023} investigated the utilisation of diffusion MRI at 0.55T to generate comprehensive maps of placental structure and function. The study involved 57 placental MRI scans acquired using a combined T2* diffusion technique capable of simultaneously capturing multiple diffusion preparations and echo times.
The results demonstrated that the quantitative parameter maps at 0.55T exhibited a high degree of similarity to those obtained at higher field strengths. This suggests that placental MRI at lower-field strengths is a viable and reliable option. 
The researchers identified several advantages of utilising lower-field strength MRI for placental imaging, including cost-effectiveness, ease of deployment, and greater patient comfort due to the wider bore.
\paragraph{}
Furthermore, Payette et al. \cite{Payette2023} investigated the potential of fetal MRI at 0.55T in conjunction with advanced image analysis techniques for clinical examination. The study introduced a pipeline for automated segmentation and quantitative analysis, addressing some of the limitations of 0.55T fetal MRI applications. 
The researchers used this imaging pipeline to acquire multi-echo dynamic sequences of the fetal body. These dynamic sequences were subsequently reconstructed into high-resolution 3D volumes, providing both structural and quantitative information.
Researchers developed a neural network model using a semi-supervised training approach to automatically segment fetal body volumes into ten different organ regions. The model demonstrated high accuracy in identifying and separating the various fetal organs from the volumetric data.
The results demonstrated a robust correlation between T2* values and gestational age in key organs, including the lungs, liver, and kidney parenchyma. It is noteworthy that the pipeline demonstrated efficacy across a broad range of gestational ages (17-40 weeks) and exhibited resilience to motion artefacts. 
The study showed that low-field fetal MRI, when integrated with advanced image analysis methods, can offer comprehensive quantitative assessments and represents a viable option for clinical scanning applications. This approach holds promise for enhancing perinatal healthcare by providing valuable insights into fetal development and monitoring.
\paragraph{}
In a study published by Silva et al. \cite{NeveSilva}, a real-time motion correction method was developed to improve functional fetal MRI scans. 
A 3D UNet was employed to track the position of the fetal brain in 125 datasets, thereby enabling real-time correction of the acquisition geometry. The method demonstrated high localisation accuracy, even for datasets of 0.55T, and was effective for fetuses under 18 weeks gestational age. This technique enhanced the reliability of fetal MRI and facilitated the characterisation of fetal brain development, with the possibility of clinical applications in conditions such as pre-eclampsia and congenital heart disease.
\paragraph{}
In 2024, the same author \cite{saraneves} developed a fully automated real-time planning system for fetal brain MRI on a 0.55T scanner to broaden the accessibility of fetal MRI. The method involved the use of DL to automatically identify crucial brain landmarks on whole-uterus echo planar imaging scans. The successful detection of these anatomical landmarks facilitated the subsequent automatic planning and guidance of radiological image acquisitions.
The pipeline was trained on over 120 datasets and subsequently evaluated for both accuracy and quality. The prospective testing of the system in nine fetal subjects demonstrated that real-time planning could be performed without any latency. 
The accuracy of landmark detection ranged from 4.2 ± 2.6 mm for the fetal eyes to 6.5 ± 3.2 mm for the cerebellum. The quality of the planned images and the diagnostic images were found to be comparable to those produced by manual planning. 
The successful implementation of real-time automatic planning has the potential to simplify fetal MRI acquisition and increase its availability in non-specialist centres.
\paragraph{}
In a recent publication, Uus et al. \cite{Uus2024} presented a novel approach to fetal MRI imaging that involves real-time 3D reconstruction of the fetal brain and body using a 0.55T MRI scanner. The researchers sought to address the fetal motion artefacts, which often compromise image quality and diagnostic accuracy. 
The method involved integrating advanced slice-to-volume reconstruction techniques into the Gadgetron framework, a platform designed for MRI data processing. By combining automated deformable and rigid reconstruction methods with DL algorithms, the research team has developed a workflow capable of processing motion-corrupted MRI data in real-time. 
This innovation not only allows for the generation of high-resolution 3D images during the scan but also enables motion correction, thereby enhancing the diagnostic utility of fetal MRI.

\begin{longtable}{|p{1.5cm}|p{4.5cm}|p{4.5cm}|p{4.5cm}|}
    \caption{Summary of Fetal Studies on Low-Field MRI Systems.}
    \label{tab6}\\
    \hline
    \textbf{Author} & \textbf{Purpose} & \textbf{Methods} & \textbf{Results} \\
    \hline
    \endfirsthead

    \multicolumn{4}{c}%
    {\tablename\ \thetable\ -- {Summary of Fetal Studies on Low-Field MRI Systems.}} \\
    \hline
    \textbf{Author} & \textbf{Purpose} & \textbf{Methods} & \textbf{Results} \\
    \hline
    \endhead

    \multicolumn{4}{r}{{(Continues on next page)}} \\
    \endfoot

    \hline
    \endlastfoot
        Ponrartana et al. (2023) \cite{Ponrartana2023} & Investigate the potential of 0.55T MRI for fetal evaluation &   0.55T MRI system with fast acquisition sequences, comprehensive fetal MRI protocol. &   High-quality fetal MRI, improved patient comfort and safety, comprehensive assessment of fetal health. \\ \hline
        Verdera et al. (2023) \cite{Verdera2023} & Assess reliability and feasibility of low-field fetal MRI for anatomical and functional measurements &   0.55T MRI scanner with a 20-minute protocol employed in 79 fetal MRI scans. &   There was concordance with control studies for anatomical measurements and high inter-observer correlations;
        
         There was a correlation between functional characteristics with gestational age. \\ \hline
        Slator et al. (2023) \cite{Slator2023} & Utilise diffusion MRI at 0.55T to generate maps of placental structure and function &   57 MRI scans using combined T2* diffusion technique. &   The maps showed high similarity to higher field strengths;
        
          0.55T is cost-effective, easy to use, and more comfortable for the patient. \\ \hline
        Payette et al. (2023) \cite{Payette2023} & Investigate the potential of 0.55T fetal MRI with advanced image analysis techniques &   Multi-echo dynamic sequences, automated segmentation and quantitative analysis pipeline. &   The method exhibits high accuracy in segmentation, correlation of T2* values with gestational age, efficacy across a broad range of gestational ages, and resilience to motion artefacts. \\ \hline
        Silva et al. \cite{NeveSilva} & Develop real-time motion correction for functional fetal MRI &   3D UNet for tracking fetal brain position in 125 datasets. &   High localisation accuracy, effective for fetuses under 18 weeks, and it has the potential to improve reliability and characterisation of fetal brain development. \\ \hline
        Saraneves et al. (2024) \cite{saraneves} & Develop fully automated real-time planning system for fetal brain MRI on a 0.55T scanner &   The objective of the DL is to identify key brain landmarks and to facilitate the automatic planning of radiological acquisitions. &   The proposed system offers real-time planning without latency, accurate landmark detection, comparable quality to manual planning, and the potential to simplify acquisition and increase availability. \\ \hline
        Uus et al. (2024) \cite{Uus2024} & Present novel approach to real-time 3D reconstruction of fetal brain and body &   The utilisation of advanced slice-to-volume reconstruction techniques, as well as the Gadgetron framework, is employed. &   The utilisation of high-resolution 3D images during the scan, in conjunction with motion correction, has the potential to enhance the diagnostic utility of fetal MRI. \\ \hline
\end{longtable}

\newpage
\section{Advances in Ultra-High-Field MRI - 7T}
\paragraph{}
There have been advances in the field of clinical MRI, especially with the FDA approval of 7T MRI in 2017. Researchers have been exploring and investigating these developments in different parts of the body in great detail.
\subsection{Brain study}
\paragraph{}
This subsection analyses studies on the application of DL in brain MRI scans. The studies explored here are shown in Table \ref{tab7}.
In 2019, Jung et al. \cite{Jung2019} developed a DL method to enhance the visualisation of perivascular spaces (PVS) in human brain MRI. 
PVS is associated with a number of brain disorders, but their quantification is challenging due to their thin and indistinct appearance. 
The proposed method utilises a complex 3D CNN with dense, hopping connections, which enables accurate prediction of the enhanced image. The networks effectively exploit contextual information from low to high-level features, resolving the issue of gradient disappearance in deep layers. 
The research team evaluated the method using 17 7T MRI images through double cross-validation. The results demonstrated that the proposed network significantly outperforms previous approaches in increasing the visibility of the PVS.
\paragraph{}
In 2020, Zong et al. \cite{Zong2020} investigated the morphology of PVS and enclosed blood vessels in healthy adults aged 21-55 years using ultra-high-resolution MRI at 7T. 
The researchers observed age-dependent changes in PVS characteristics, including an increased count and volume fraction in the basal ganglia, as well as spatial heterogeneity. The use of carbogen breathing induced dynamic alterations in PVS morphology, indicating that this system is sensitive to physiological changes. These results provide valuable insights into the glymphatic system and its potential implications for brain health.
\paragraph{}
Meliadò et al. \cite{Meliadò} introduced a DL approach for image-based subject-specific assessment of local SAR, which is crucial for ensuring patient safety during MRI procedures, especially at 7T. 
In the past, SAR assessment has been conducted using offline numerical simulations with generic body models, which has led to uncertainties due to inter-subject anatomical variability. 
The proposed method employs a CNN trained on subject-specific maps and corresponding local SAR distributions, thereby enabling more accurate and efficient SAR estimation. By incorporating additional penalties for SAR underestimation errors in the loss function, their model achieves remarkable results in both \textit{in silico} and \textit{in vivo} validations, demonstrating excellent qualitative and quantitative agreement with ground truth data. 
This approach not only enhances the accuracy of SAR assessment but also reduces examination protocol time by nearly 25\%, thus improving patient safety and MRI efficiency in clinical practice.
\paragraph{}
Yang et al. \cite{Yang2020} introduced the Automated Hippocampal Subfield Segmentation Toolbox (CAST), a deep CNN designed for the segmentation of hippocampal subfields in 7T and 3T MRI datasets. 
The CAST model is adaptable, enabling researchers to train it with their data and accommodating various modalities. 
The results demonstrated that CAST exhibited comparable segmentation accuracy to a standard method on the 7T and 3T datasets. The CAST method reliably segments subfields, prioritising efficiency and accuracy, particularly in 7T MRI applications. The toolbox employs a multi-scale deep CNN to capture global structural information while efficiently using computational resources. 
Therefore, CAST markedly enhances the reliability of subfield segmentation and ensures consistent segmentation quality across diverse datasets.
\paragraph{}
Bazin et al. \cite{Bazin2020} developed a method for measuring image sharpness in 7T MRI images. The aim of the study was to investigate the association between subject movement and image sharpness, with a particular focus on studies of brain structure. 
The method was assessed in 24 healthy volunteers, including individuals of varying ages. The results demonstrated a correlation between motion and sharpness enhancement, with motion correction exhibiting a pronounced impact on sharpness. 
Following motion correction, the full width at half maximum was decreased from 0.88 mm to 0.70 mm, resulting in a 2.0 times smaller voxel volume. 
This method provides valuable insights into the impact of motion on image quality in 7T MRI studies.
\paragraph{}
Kim et al. \cite{Kim2020} presented DCN-Net, a DL framework designed for the accurate and rapid segmentation of cerebellar dentate nuclei in 7T diffusion MRI. 
The DCN-Net employs a hybrid loss function to address the issue of highly imbalanced data, and it learns the probability of each label independently during training, including those for dentate and intermediate nuclei. 
To address the limited availability of labelled data, auto-training strategies are employed. These strategies generate auxiliary labels for unlabelled data using DCN-Net, which has been trained with manual labels. 
The experimental results, based on 60 individuals, demonstrated that DCN-Net outperforms other segmentation tools and deep neural networks in terms of accuracy and consistency.
\paragraph{}
In 2021, Demirel et al. \cite{Demirel2021} presented a new method for enhancing 7T MRI technology, achieving a 20-fold acceleration in imaging while maintaining high image quality. 
This development is important for enhancing spatial and temporal resolution in brain imaging, and for accurately mapping neural activity. The researchers addressed the limitations of traditional techniques, which suffer from aliasing and noise at higher rates. 
The study demonstrates significant improvements over existing methods by employing a self-supervised DL reconstruction method that uses only undersampled datasets. This offers comparable functional precision and temporal effects to standard 10-fold accelerations.
\paragraph{}
Trutti et al. \cite{Trutti2021} presented a probabilistic atlas of the human ventral tegmental area (VTA) using 7T MRI data, addressing the need for anatomically atlases of subcortical structures. 
While traditional neuroanatomical atlases are invaluable for localising brain regions using functional MRI, they often lack coverage of subcortical nuclei, limiting their utility for deep brain imaging. 
The researchers developed a multimodal 7T MRI protocol for enhanced imaging and delineation of the VTA, leveraging ultra-high field resolution. 
This was followed by the creation of a comprehensive probabilistic VTA atlas.
\paragraph{}
In a subsequent study, Morris et al. \cite{Morris2022} utilised high-resolution 7T MRI to investigate the integrity of the VTA in individuals with and without depression and anxiety disorders.
By employing quantitative MRI and dopamine-related signal mapping, they were able to delineate the VTA and correlate its structural integrity with measures of intrinsic and extrinsic motivation assessed through cognitive tasks. 
The results demonstrated that subjects with mood and anxiety disorders exhibited larger VTA volumes but lower signal intensity within the VTA compared to healthy controls. This indicates reduced structural integrity of the dopaminergic VTA in these individuals. 
It should be noted that VTA integrity was not directly associated with self-reported symptoms of depression or anxiety. However, it did show a correlation with objective measures of extrinsic motivation, suggesting a link between neural integrity and behaviour in clinical and non-clinical groups.
\paragraph{}
In 2022, Miletić et al. \cite{Miletić} conducted a study using 7T functional MRI to investigate the tripartite model of the subthalamic nucleus during a speeded decision-making task. 
The tripartite model, which suggests cognitive, limbic, and motor subregions within the subthalamic nucleus, is typically supported by anatomical studies and research on clinical patients. 
In this study, 34 healthy participants were asked to complete a random-dot motion task with varying levels of difficulty and choice payoffs. The objective was to differentially engage the cognitive and limbic networks while recording motor responses using the left and right index fingers. The analysis of the signals in manually delineated subthalamic nucleus subregions revealed that all segments responded similarly to the experimental manipulations. 
The study found no evidence to support the existence of distinct functional subregions, challenging the tripartite model by demonstrating uniform activity across the subthalamic nucleus subregions in response to the task.
\paragraph{}
Rodrigues et al. \cite{Rodrigues2022} proposed a benchmark for hypothalamus segmentation on 7T MRI images, addressing the challenges of high variability and subjectivity in manual segmentation due to unclear morphological landmarks. 
The study presents a diverse dataset of 1381 individuals from various sources.
The researchers proposed a teacher-student model with segmentation and correction blocks, which they demonstrated to have improved generalisation ability. The method was trained on the datasets, achieving Dice coefficients of 0.83 on familiar datasets and 0.74 on the unseen dataset. 
This demonstrates the effectiveness and reliability of the approach, offering a valuable tool for research in automated hypothalamus segmentation on ultra-high-resolution 7T MRI images.
\paragraph{}
Wei et al. \cite{Wei2022} developed CaNes-Net, a cascaded nested network designed for segmenting 3T MRI images. The authors addressed the issue of low tissue contrast and anatomical variability in 3T images by utilising nested networks that iteratively refine the segmentation. The network is trained using the higher contrast of 7T MRI images. 
To address the issue of misalignment between 3T and 7T images, the researchers have incorporated a correlation coefficient map. This map enables the segmentation accuracy to be improved, with results that are superior to those achieved by popular tools. 
The CaNes-Net is constituted of two nested networks: The Nes-Net handles the initial segmentation, while another Nes-Net refines the segmentation using tissue-specific distance maps. This iterative process allows for the gradual improvement of the segmentation results, ensuring optimal results are achieved. 
The CaNes-Net was tested on 18 adults and the ADNI dataset, demonstrating reduced segmentation errors and superior performance compared to widely used methods. 
This highlights the effectiveness of incorporating 7T images in the segmentation process.
\paragraph{}
In 2023, Doss et al. \cite{Doss2023} conducted a study to develop a DL network for accurately segmenting the basal nucleus of Meynert in 3T MRI scans. This structure is important for cognitive function and arousal.
The small size, patient variability, and low contrast on 3T MRI make accurately segmenting this structure a challenging task. 
The researchers employed a dataset including 21 healthy individuals, each with paired 3T and 7T MRI scans. The 7T scans were annotated, resulting in accurate labels for the 3T images. A 3D-Unet CNN was constructed and validated using a 5-fold cross-validation approach. 
The model demonstrated superior performance compared to the standard probabilistic atlas, with higher Dice coefficients and reduced centroid distances observed in both healthy subjects and patients with temporal lobe epilepsy.
\paragraph{}
Zhang et al. \cite{Zhang2022} developed a model, the quantitative Recurrent Inference Machine (qRIM), which addresses the challenges of long scan times in 7T MRI. 
This model employs DL techniques to reconstruct images and map values from sparse data. Integrated into a Recurrent Inference Machine, qRIM solves the inverse problem iteratively, making use of data redundancy and prior information from the relaxometry model. 
Experimental evaluations using ultra-high-resolution multi-echo gradient echo data have demonstrated that qRIM markedly reduces reconstruction error in comparison to sequential reconstruction and parameter fitting methods. 
Moreover, it outperforms other architectures, such as U-Net and Compressed Sensing, and can be used to investigate human subcortex activities across all age groups.
Although the quantitative end-to-end variational network exhibits slightly inferior reconstruction quality in comparison to qRIM, it nevertheless outperforms alternative methods. 
In general, qRIM demonstrates superior performance with increasing acceleration factors and benefits the reconstruction quality of sparser data.
\paragraph{}
Quantitative susceptibility mapping (QSM) is a non-invasive MRI technique that makes it possible to measure the spatial distribution of magnetic susceptibility in an object. 
Chen et al. \cite{Chen2022} evaluated various QSM methods to assess radiation-induced cerebral microbleeds and basal ganglia at both 3T and 7T MRI field strengths. 
The study tested four background phase removal methods and five dipole field inversion algorithms, finding that 7T MRI provided a lower noise level compared to 3T. 
The QSIP (Quantitative Susceptibility mapping by Inversion of a Perturbation) and VSHARP (Sophisticated Harmonic Artifact Reduction for Phase) + iLSQR (iterative Least Squares with QR factorisation) methods yielded the best results for white matter homogeneity and vein contrast, with QSIP also giving the highest cerebral microbleeds contrast.
The results demonstrated that VSHARP + iLSQR produced the most accurate susceptibility values across all regions, closely matching the ground truth. 
This research provides guidance on selecting the most appropriate QSM processing pipeline for specific applications and MRI field strengths.
\paragraph{}
In 2023, Ramadass et al. \cite{Ramadass2023} investigated the challenges and solutions to achieving accurate segmentation of the whole brain using 7T structural MRI.
The study showed the potential of 7T MRI to enhance the comprehension of brain function due to its superior resolution and contrast. However, existing segmentation tools encounter difficulties when confronted with inhomogeneous image contrast on 7T MRI. 
To address this issue, the researchers proposed a solution: registering the segmentation of the whole brain in 3T to 7T MRI and adjusting the model accordingly. This approach led to a significant reduction in the average relative difference in label volumes and an improvement in the Dice similarity coefficient, indicating a more accurate segmentation. 
The results demonstrate that fine-tuning the model is an effective method for improving whole-brain segmentation in 7T MRI, thereby enabling more detailed and accurate neuroimaging analyses.
\paragraph{}
Arledge et al. \cite{Arledge2022} developed a DL approach to automatically generate vascular pharmacokinetic parameter maps from dynamic contrast-enhanced MRI images in mouse brain tumour models. 
The approach was validated using 7T MRI scans performed on glioma in nude mice. The researchers trained a CNN on dynamic contrast-enhanced images to capture multiscale features. 
The CNN demonstrated promising results, accurately estimating vascular pharmacokinetic parameters and matching conventional pharmacokinetic models. 
Furthermore, the DL approach was successfully applied to a mouse model of breast cancer brain metastases, indicating its potential for assessing vascular permeability in different types of brain tumours.
\paragraph{}
Brink et al. \cite{Brink2022} developed a method for personalising local SAR predictions for parallel transmit (PTx) neuroimaging at 7T, utilising a T1-weighted dataset. 
The researchers addressed the challenge of regulatory compliance and RF performance limitations presented by intersubject variability in SAR. By employing DL, they generated subject-specific body models, thereby overcoming the limitations of the conventional "one-size-fits-all" approach. 
The CNN developed by the researchers trained on multi-contrast 7T MRI data, achieved accurate segmentations, addressing the intensity bias and contrast variations that are typical at 7T.
The efficacy of personalised local SAR predictions was demonstrated through evaluation in PTx configurations.
\paragraph{}
In a subsequent study, Gokyar et al. \cite{Gokyar2023} presented a DL-based method for predicting SAR distributions in the human head for 7T MRI PTx systems. At 7T, the high intensity of the magnetic field results in a non-uniform deposition of electromagnetic energy, which can lead to the formation of local high SAR hot spots.
To address this issue, the researchers developed a multichannel 3D CNN designed to predict local SAR maps and peak spatial SAR levels. The CNN architecture demonstrated superior prediction accuracy, with an average reduction in overestimation errors of 20\% compared to 152\% with traditional methods. 
The DL model demonstrated a reduction in prediction uncertainties and was able to perform predictions in a relatively short time.
\paragraph{}
In their study, Liu et al. \cite{Liu2022} proposed a new unsupervised learning method, self-labelled encoder-decoder (SLED), to enhance the reconstruction of myelin water image (MWI) maps from multi-echo gradient echo sequences. 
The objective of this study was to address the challenges associated with stability, accuracy, and SNR in adapting MWI data. 
The imaging data was obtained utilising a 7T MRI system designed for preclinical research applications. The implementation of the SLED network enabled the researchers to demonstrate an improvement in the estimation of myelin water fraction (MWF) compared to the traditional approach. 
Experiments conducted on ultra-high resolution MWI data from rat brains and a simulated MWI phantom demonstrated that SLED exhibited superior stability and accuracy in MWF estimation. 
The generated maps exhibited less noise and demonstrated greater tolerance to low SNR data. It is of particular significance that SLED's unsupervised, self-labelling approach represents a distinctive alternative for the analysis of MWI.
\paragraph{}
In 2023, Tsuji et al. \cite{Tsuji2023} investigated the use of DL to enhance image quality and reduce acquisition time in preclinical 7T MRI, focusing on brain images of rats. 
The researchers employed a residual dense network (RDN) to reconstruct images with a low number of excitations (NEXs), to match the quality of high NEX images while reducing scan time. 
By training the RDN with low NEX images as input and high NEX images as the reference, they achieved significant improvements in SNR, peak SNR, and SSIM. 
The results demonstrated that the enhanced images produced by the RDN exhibited higher quality, with the SNR for NEX = 2 increasing from 10.4 to 32.1, surpassing even the high NEX (NEX = 12) SNR of 19.6. 
Furthermore, the method reduced the scanning time by 83\%, demonstrating its potential for efficient and high-quality MRI imaging.
\paragraph{}
In a more recent study, Jha et al. \cite{Jha2023} examined the diffusion MRI and its application in the assessment of white matter regions of the brain. The researchers proposed a novel approach to transform diffusion MRI data acquired at 3T to achieve a quality comparable to that obtained at 7T. 
This is achieved by combining the trapezoidal rule with graph-based attention modules. 
This transformation is of significant importance, as higher magnetic field strengths, such as 7T, afford greater tissue contrast and spatial resolution, thereby enabling more accurate estimation of structural details within the brain. 
However, due to the cost of 7T scanners, there is a pressing need for algorithms to enhance the quality of 3T diffusion MRI data. 
The proposed CNN architecture was designed to resolve this limitation and the results confirm that it is a viable alternative.
\paragraph{}
An et al. \cite{An2023} developed a DL-based method for fully automated segmentation of ischemic stroke lesions in 7T MRI data acquired from mouse models.
This approach addresses the labour-intensive and variable nature of manual segmentation by trained experts. 
Utilising a dataset of 382 MRI scans, the researchers split the data into 293 scans for training and 89 for evaluation. 
The evaluation of the automated segmentation technique involved comparing its output with manual segmentation by calculating Dice coefficients and assessing the accuracy of lesion volume measurements. Additionally, the automated method was further tested on an independent dataset with different imaging characteristics to assess its robustness.
The automated segmentation technique produced segmentation masks that appeared smooth, compact, and realistic, achieving a high level of agreement with the manual segmentation performed by human experts.
Remarkably, the Dice scores surpassed those between human experts in previous studies, indicating the method's potential to eliminate human bias and standardise lesion segmentation across various research studies and centres.

\begin{longtable}{|p{1.5cm}|p{4.5cm}|p{4.5cm}|p{4.5cm}|}
    \caption{Summary of Brain Studies on Ultra-High-Field MRI Systems.}
    \label{tab7}\\
    \hline
    \textbf{Author} & \textbf{Purpose} & \textbf{Methods} & \textbf{Results} \\
    \hline
    \endfirsthead

    \multicolumn{4}{c}%
    {\tablename\ \thetable\ -- {Summary of Brain Studies on Ultra-High-Field MRI Systems.}} \\
    \hline
    \textbf{Author} & \textbf{Purpose} & \textbf{Methods} & \textbf{Results} \\
    \hline
    \endhead

    \multicolumn{4}{r}{{(Continues on next page)}} \\
    \endfoot

    \hline
    \endlastfoot

    Jung et al. (2019) \cite{Jung2019} & Enhance visualisation of PVS in brain MRI &   Developed a 3D CNN with dense, hopping connections. &   Improved visibility and quantification of PVS, outperforming previous methods. \\ \hline
    Zong et al. (2020) \cite{Zong2020} & Investigate morphology of PVS and blood vessels in healthy adults &   Used ultra-high-resolution 7T MRI to study age-dependent changes. &   Found increased PVS count and volume with age, and sensitivity to physiological changes. \\ \hline
    Meliadò et al. (2020) \cite{Meliadò} & Subject-specific assessment of local SAR in MRI &   Employed a CNN trained on subject-specific maps and SAR distributions. &   Achieved accurate SAR estimation, reducing examination time by 25\%. \\ \hline
    Yang et al. (2020) \cite{Yang2020} & Segmentation of hippocampal subfields in MRI datasets &   Developed the CAST deep CNN adaptable for 7T and 3T MRI data. &   Achieved comparable segmentation accuracy to state-of-the-art methods. \\ \hline
    Bazin et al. (2020) \cite{Bazin2020} & Measure image sharpness in 7T MRI images &   Evaluated the relationship between subject movement and image sharpness. &   Demonstrated significant impact of motion correction on image quality. \\ \hline
    Kim et al. (2020) \cite{Kim2020} & Segmentation of cerebellar dentate nuclei in 7T diffusion MRI &   Developed DCN-Net with a hybrid loss function and auto-training strategies. &   Outperformed other segmentation tools in accuracy and consistency. \\ \hline
    Demirel et al. (2021) \cite{Demirel2021} & Enhance 7T MRI imaging speed while maintaining quality &   Used a self-supervised DL reconstruction method with undersampled datasets. &   Achieved 20-fold acceleration with comparable functional precision. \\ \hline
    Trutti et al. (2021) \cite{Trutti2021} & Develop a probabilistic atlas of the human VTA &   Developed a comprehensive probabilistic atlas for subcortical structures. &   Created a protocol for multimodal VTA imaging and delineation using 7T MRI. \\ \hline
    Morris et al. (2022) \cite{Morris2022} & Investigate VTA integrity in subjects with and without mood disorders &   Used 7T MRI to correlate VTA structure with cognitive measures. &   Found larger VTA volumes and lower signal intensity in subjects with mood disorders. \\ \hline
    Miletić et al. (2022) \cite{Miletić} & Investigate subthalamic nucleus model during decision-making tasks &   Used 7T fMRI with a speeded decision-making task. &   Found uniform activity across subthalamic nucleus subregions, challenging existing models. \\ \hline
    Rodrigues et al. (2022) \cite{Rodrigues2022} & Benchmark hypothalamus segmentation on 7T MRI images &   Developed a teacher-student model with segmentation and correction blocks. &   Achieved high Dice coefficients, demonstrating effective segmentation. \\ \hline
    Wei et al. (2022) \cite{Wei2022} & Segment 3T MRI images using 7T MRI data &   Developed CaNes-Net with nested networks and correlation coefficient maps. &   Improved segmentation accuracy, outperforming popular tools. \\ \hline
    Doss et al. (2023) \cite{Doss2023} & Segment basal nucleus of Meynert in 3T MRI scans &   Constructed and validated a 3D-Unet CNN using 5-fold cross-validation. &   Demonstrated superior performance compared to standard atlas methods. \\ \hline
    Zhang et al. (2022) \cite{Zhang2022} & Reduce scan times in 7T MRI &   Developed qRIM to reconstruct images and map values from sparse data. &   Reduced reconstruction error, outperforming other methods. \\ \hline
    Chen et al. (2022) \cite{Chen2022} & Assess radiation-induced CMB and basal ganglia &   Several QSM methods were subjected to testing at both 3T and 7T. &   VSHARP + iLSQR is the most accurate method, while QSIP is the best for microbleeds contrast. \\ \hline
    Ramadass et al. (2023) \cite{Ramadass2023} & Improve whole-brain segmentation in 7T MRI &   Registered 3T segmentation to 7T MRI and adjusted the model. &   Achieved significant reduction in segmentation errors. \\ \hline
    Arledge et al. (2022) \cite{Arledge2022} & Generate vascular pharmacokinetic parameter maps in mouse brain tumours &   Trained a CNN on dynamic contrast-enhanced images using 7T MRI. &   Accurately estimated vascular parameters, applicable to various brain tumors. \\ \hline
    Brink et al. (2022) \cite{Brink2022} & Personalise local SAR predictions for 7T neuroimaging &   Developed a CNN using multi-contrast 7T data for subject-specific body models. &   Demonstrated accurate segmentations and personalised SAR predictions. \\ \hline
    Gokyar et al. (2023) \cite{Gokyar2023} & Predict SAR distributions in 7T MRI PTx systems &   Developed a multichannel 3D CNN for local SAR maps prediction. &   Achieved superior prediction accuracy, reducing overestimation errors. \\ \hline
    Liu et al. (2022) \cite{Liu2022} & Enhance reconstruction of myelin water image maps at 7T MRI &   Developed SLED network for multi-echo gradient echo sequences. &   Improved stability and accuracy in myelin water fraction estimation. \\ \hline
    Tsuji et al. (2023) \cite{Tsuji2023} & Improve image quality and reduce acquisition time in preclinical 7T MRI &   Used a residual dense network to reconstruct low NEX images. &   Achieved higher quality images with reduced scan time. \\ \hline
    Jha et al. (2023) \cite{Jha2023} & Transform diffusion MRI data acquired at 3T to mimic 7T quality &   The methodology employed in this study includes the trapezoidal rule, graph-based attention modules and a CNN architecture. &   Improvements in tissue contrast and spatial resolution, accurate estimation of structural details, and enhanced 3T diffusion MRI data quality. \\ \hline
    An et al. (2023) \cite{An2023} & Segment ischemic stroke lesions in mouse 7T MRI &   Utilised a CNN trained on T2-weighted MRI data for automated segmentation. &   Achieved high agreement with manual segmentation, surpassing human raters. \\ \hline
\end{longtable}

\subsubsection{Neurological Diseases}
\paragraph{}
This subsection analyses studies on the application of DL in brain MRI scans relating to diseases of the central nervous system. The studies presented here are shown in Table \ref{tab8}.
In 2019, Shamir et al. \cite{Shamir2019} evaluated the accuracy of visualising the subthalamic nucleus using 7T MRI combined with ML for deep brain stimulation (DBS) in Parkinson's disease patients. 
They developed 7T-ML, a method that uses a database of 7T MRIs from Parkinson's patients and ML algorithms to visualise the subthalamic nucleus in standard clinical MRI images.
The results showed high agreement between the 7T-ML method and intraoperative microelectrode recordings, indicating its reliability for patient-specific guidance of the subthalamic nucleus during DBS surgery.
\paragraph{}
In their study, La Rosa et al. \cite{LaRosa} introduced CLAIMS, an AI-based framework for the automatic detection and classification of cortical lesions in multiple sclerosis (MS) using 7T MRI. 
CLAIMS represents a significant improvement upon existing manual segmentation methods, which are constrained by time limitations and moderate reliability. 
The results demonstrated high accuracy, particularly in the identification of different types of cortical lesions. Furthermore, the technique outperformed existing techniques, indicating potential for improving clinical decision-making in MS diagnosis.
\paragraph{}
Barry et al. \cite{Barry2021} employed 7T functional MRI to investigate neurodegenerative changes in amyotrophic lateral sclerosis (ALS). The aim was to identify non-invasive biomarkers to localise, measure and monitor ALS processes. The study involved 12 ALS participants and nine age-matched controls who were scanned 7T functional MRI during a resting state. 
The analysis revealed disruption of long-range functional connectivity between the superior sensorimotor cortex and bilateral cerebellar lobe VI in ALS participants, a region associated with complex motor and cognitive processing.
The results contribute evidence of cerebellar involvement in ALS pathology.
\paragraph{}
In 2022, Lancione et al.\cite{Lancione2022} investigated the potential of QSM in conjunction with 7T MRI to enhance the diagnostic accuracy of multiple system atrophy (MSA). 
The study examined the influence of echo time on susceptibility values and evaluated the efficacy of histogram analysis in enhancing diagnostic accuracy. The study included 32 patients diagnosed with MSA and 16 healthy controls. MRI scans were employed to generate susceptibility maps, from which histogram features were extracted and compared.
MSA is characterised by elevated iron burden in early-affected subcortical nuclei.
The study revealed significant alterations in susceptibility in specific brain regions, with increased iron deposition observed in the putamen, substantia nigra, globus pallidus, caudate nucleus, and dentate nucleus for MSA. 
It was found that shorter echo time were more effective in capturing these changes, with histogram features demonstrating high diagnostic accuracy, as reflected in curve areas exceeding 0.9. Therefore, the study showed that the use of short echo time and histogram analysis significantly improves the diagnostic potential of QSM in the detection of subtle variations in iron deposition in MSA.
\paragraph{}
Donnay et al. \cite{Donnay2023} developed a method for automatic segmentation of 7T MRI scans in multiple sclerosis (MS) patients, addressing challenges such as bias fields, susceptibility artefacts, and registration errors. 
The researchers introduced Pseudo-Label Assisted nnU-Net (PLAn), a DL approach that combines removal of the skull and segmentation of the entire brain. 
The study involved two cohorts: Cohort 1 included 25 MS patients with both 3T and 7T scans, while Cohort 2 had 8 MS patients scanned only at 7T. 
PLAn was trained and optimised using Cohort 1 data and validated on Cohort 2. 
The results demonstrated that PLAn outperformed other methods in lesion detection, achieving a 16\% improvement in the Dice Similarity Coefficient compared to nnU-Net. 
Consequently, the efficacy of PLAn in accurately segmenting high-field MRI scans, particularly in MS patients, is evident.
\paragraph{}
In a comparative study on various QSM methods for iron-sensitive imaging at 7T, Yao et al. \cite{Yao2023} focused on healthy individuals and patients with Huntington's disease (HD). The study evaluated several dipole inversion algorithms, including iterative least-squares, single-step methods, and DL-based techniques. 
The performance of these methods was evaluated in terms of their deviation from a multiple orientation QSM reference, visual quality, susceptibility measurement in subcortical regions, correlation with postmortem iron quantification, and their ability to differentiate between HD patients and healthy controls. 
The single-step QSM methods with total variation or total generalised variation constraints, along with the single-step DL method iQSM, demonstrated the strongest correlation with iron deposition and were most effective in distinguishing premanifest HD subjects from healthy controls. 
The DL method trained with multiple orientation data produced QSM maps that were most similar to the multiple orientation reference and had the best visual scores.
Therefore, the study has made significant progress in the accuracy and effectiveness of QSM methods, with important implications for the diagnosis and monitoring of Huntington's disease.
\paragraph{}
In his most recent article, Jakary et al. \cite{Jakary2023} used a 7T phase sensitive neuroimaging to investigate the influence of brain iron levels on depression severity and cognitive function in individuals undergoing mindfulness-based cognitive therapy (MBCT) for major depressive disorder (MDD). 
17 unmedicated MDD participants and 14 healthy controls (HC) underwent MRI, depression severity evaluation, and cognitive testing before and after MBCT. Local field shift (LFS) values, indicating brain iron levels, were derived from phase images in specific brain regions. The MDD group exhibited lower baseline LFS in certain regions compared to HC and demonstrated impairment in information processing speed. Lower LFS values in various brain regions were associated with depression severity and cognitive performance. MBCT led to depression relief and improved executive function and attention. 
Participants with specific baseline LFS values experienced greater improvement in depression severity and cognitive function post-MBCT. The study emphasises the potential role of brain iron in the symptoms of MDD and the effectiveness of its treatment.

\begin{longtable}{|p{1.5cm}|p{4.5cm}|p{4.5cm}|p{4.5cm}|}
    \caption{Summary of Neurological Diseases Studies on Ultra-High-Field MRI Systems.}
    \label{tab8}\\
    \hline
    \textbf{Author} & \textbf{Purpose} & \textbf{Methods} & \textbf{Results} \\
    \hline
    \endfirsthead

    \multicolumn{4}{c}%
    {\tablename\ \thetable\ -- {Summary of Neurological Diseases Studies on Ultra-High-Field MRI Systems.}} \\
    \hline
    \textbf{Author} & \textbf{Purpose} & \textbf{Methods} & \textbf{Results} \\
    \hline
    \endhead

    \multicolumn{4}{r}{{(Continues on next page)}} \\
    \endfoot

    \hline
    \endlastfoot
        Shamir et al. (2019) \cite{Shamir2019} & Evaluate visualisation accuracy of the subthalamic nucleus using 7T MRI combined with ML for DBS in PD &   The 7T-ML method employs a database of 7T MRI scans and ML algorithms. &   It has been demonstrated that there is a high degree of agreement between the 7T-ML method and intraoperative microelectrode recordings. \\ \hline
        La Rosa et al. (2020) \cite{LaRosa}& Automatic detection and classification of cortical lesions in MS using 7T MRI &   Develop the CLAIMS AI-based framework. &   The technique offers high accuracy in identifying cortical lesions, outperforming existing techniques. \\ \hline
        Barry et al. (2021) \cite{Barry2021} & Investigate neurodegenerative changes in ALS using 7T fMRI &   The study employs high-resolution resting-state fMRI at 7T. &   The disruption of long-range functional connectivity suggests that the cerebellum is involved in ALS. \\ \hline
        Lancione et al. (2022) \cite{Lancione2022}& Enhance diagnostic accuracy of MSA using QSM with 7T MRI &   A histogram analysis of the susceptibility maps was conducted. &   Significant alterations in susceptibility, increased iron deposition, and high diagnostic accuracy. \\ \hline
        Donnay et al. (2023) \cite{Donnay2023} & Automatic segmentation of 7T MRI scans in MS patients &   A Pseudo-Label Assisted nnU-Net (PLAn) was developed. &   The PLAn demonstrated superior performance in lesion detection, with a 16\% improvement in the Dice Similarity Coefficient compared to nnU-Net. \\ \hline
        Yao et al. (2023) \cite{Yao2023}& Evaluate various QSM methods for iron-sensitive imaging in HD patients and healthy subjects &   A variety of dipole inversion algorithms, including those based on the DL approach, were employed. &   The single-step QSM methods and DL method iQSM demonstrated the strongest correlation with iron deposition, with the capacity to effectively distinguish HD subjects from healthy controls. \\ \hline
        Jakary et al. (2023) \cite{Jakary2023}& Investigate the influence of brain iron levels on depression severity and cognitive function in MDD patients undergoing MBCT &   The neuroimaging technique employed was 7T phase-sensitive imaging, with the LFS values derived from the phase images. &   A lower baseline LFS was associated with greater depression severity and poorer cognitive performance;

          MBCT led to improvement. \\ \hline
\end{longtable}

\newpage
\subsection{Musculoskeletal Study}
\paragraph{}
This subsection examines studies on the application of DL in muscle and skeletal MRI scans. The studies presented here are shown in Table \ref{tab9}.
In 2021, Belyk et al. \cite{Belyk2021} conducted a study investigating the coordination of vocal and respiratory control in the human brain. 
The researchers focused on the larynx motor cortices involved in vocal-motor control and employed a 7T functional MRI to localise and understand the function of these cortical regions. The study involved 13 participants who were required to complete a series of tasks involving singing and whistling. 
In contrast to previous assumptions, the study found that both the dorsal and ventral larynx motor regions are activated during whistling, despite the decreased engagement of the larynx. This indicates that both areas exhibit a shared property of laryngeal-respiratory integration. 
\paragraph{}
Tanaka et al. \cite{Tanaka2022} sought to discern whether the initial reduce in the blood oxygenation level-dependent signal contains information that can be decoded to identify facial expressions. 
The researchers employed multivoxel pattern analysis and hyperalignment techniques on 7T functional MRI data to analyse brain activity during a facial expression discrimination task. 
The results demonstrate that the decoding accuracy in the bilateral amygdala exceeds chance levels 2 seconds after face onset. This indicates that the initial reduction in blood oxygenation level-dependent signals contains more precise information about task events and cognitive processes.
\paragraph{}
In 2022, Terraciano et al. \cite{Terracciano2022} developed a 7T MRI protocol using delayed gadolinium-enhanced MRI of cartilage to quantitatively assess post-traumatic osteoarthritis (PTOA) in a rabbit model. 
Following unilateral anterior cruciate ligament (ACL) transactions in rabbits, MRI examinations were performed at various intervals using a modified delayed gadolinium-enhanced MRI of cartilage protocol to study PTOA development. 
Voxel-based maps were generated to evaluate the changes in cartilage. 
The study demonstrated a decline in the delayed gadolinium-enhanced index in the medial compartment of the knee following ACLT, indicative of cartilage degeneration. 
Histological analysis confirmed the presence of PTOA in ACLT knees, with osteophytes observed exclusively in these knees. This approach demonstrates the feasibility of detecting and monitoring PTOA using quantitative 7T MRI, which has the potential to facilitate therapeutic interventions and advance our understanding of PTOA progression.

\begin{longtable}{|p{1.5cm}|p{4.5cm}|p{4.5cm}|p{4.5cm}|}
    \caption{Summary of Musculoskeletal Studies on Ultra-High-Field MRI Systems.}
    \label{tab9}\\
    \hline
    \textbf{Author} & \textbf{Purpose} & \textbf{Methods} & \textbf{Results} \\
    \hline
    \endfirsthead

    \multicolumn{4}{c}%
    {\tablename\ \thetable\ -- {Summary of Musculoskeletal Studies on Ultra-High-Field MRI Systems.}} \\
    \hline
    \textbf{Author} & \textbf{Purpose} & \textbf{Methods} & \textbf{Results} \\
    \hline
    \endhead

    \multicolumn{4}{r}{{(Continues on next page)}} \\
    \endfoot

    \hline
    \endlastfoot
        Belyk et al. (2021) \cite{Belyk2021} & Investigate coordination of vocal and respiratory control in the brain &   7T fMRI, tasks involving singing and whistling with 13 participants. &   Both dorsal and ventral larynx motor areas activated during whistling, indicating laryngeal-respiratory integration. \\ \hline
        Tanaka et al. (2022) \cite{Tanaka2022} & Decode facial expressions from an initial decrease in BOLD signal &   Multi-voxel pattern analysis, hyper-alignment techniques on 7T fMRI data. &   Decoding accuracy in the bilateral amygdala exceeds chance levels, providing insights into neural mechanisms of facial expression recognition. \\ \hline
        Terraciano et al. (2022) \cite{Terracciano2022} & Assess PTOA in a rabbit model &   7T MRI with delayed gadolinium enhancement protocol, ACL transactions in rabbits. &   Decrease in gadolinium enhancement index in knee cartilage post-ACLT, histological analysis confirmed the presence of PTOA. \\ \hline
\end{longtable}

\subsection{B0 Inhomogeneity study}
\paragraph{}
This subsection analyses studies that aim to solve the problem of the inhomogeneity of the MRI magnetic field. The studies presented here are shown in Table \ref{tab10}.
In a study published in 2021, Haast et al. \cite{Haast2021} examined the impact of MP2RAGE B1+ sensitivity on T1 reproducibility and hippocampal morphometry in 7T MRI. 
MP2RAGE stands for Magnetization-Prepared 2 RApid Gradient Echo. The sequence is designed to provide high-resolution anatomical images with improved contrast and reduced bias from factors such as magnetic field inhomogeneities and transmit field (B1+) variations. 
The study, conducted across multiple sites, aimed to validate the effects of specific MP2RAGE parameter choices on image quality and cortical T1 estimates, while also evaluating their consequences on hippocampal segmentation results. 
The results of the study underline the importance of optimising MP2RAGE sequences for neuroimaging studies, particularly in the context of 7T MRI, highlighting the need for standardisation and harmonisation of MRI protocols at different imaging sites.
\paragraph{}
In 2022, Ma et al. \cite{Ma2022} introduced a DL approach aimed at mitigating transmit-B1 (B1+) artefacts in MRI without relying on PTx hardware. 
The method involved training a deep neural network to predict PTx-style images, and it demonstrated promising results in improving diffusion tensor imaging analysis. This was achieved by effectively restoring signal dropout and improving image quality. 
Furthermore, the trained DL model was able to successfully predict PTx images for new subjects, thereby demonstrating its potential to reduce B1+ artefacts even in the absence of PTx resources.
\paragraph{}
Hunger et al. \cite{Hunger2023} developed a DL approach, named deepCEST 7T, to enhance the speed and precision of chemical exchange saturation transfer MRI parameter mapping at 7T. 
This method employs a neural network to address the pronounced B1 inhomogeneity challenges that are characteristic of 7T. 
The network is trained to predict parameters from a single scan using 7T \textit{in vivo} and B1 maps as input. This process results in a reduction in scan time to 6 minutes and 42 seconds while maintaining the ability to correct for both B0 and B1 inhomogeneities.
Xu et al. \cite{Xu2023} presented the autoHOS, an automated high-order shimming method tailored for addressing B0 inhomogeneity in MRI and MR spectroscopy at 7T. 
This approach combines DL based brain extraction with image-based high-order shimming, eliminating manual ROI selection. 
\textit{In vivo} studies at 7T have demonstrated that the autoHOS method enhances image and spectral quality in comparison with the traditional linear and manual shimming methods.
\paragraph{}
Zhao et al. \cite{Zhao2023} developed a distortion-free whole-cerebrum 3D pseudo-continuous arterial spin labelling (pCASL) sequence at 7T to overcome B1/B0 inhomogeneities. 
By optimising pCASL labelling parameters, background suppression pulses, and an accelerated Turbo-FLASH readout, they achieved high-resolution pCASL imaging with whole-brain coverage and detailed perfusion information. 
The proposed technique demonstrated excellent test-retest repeatability and improved SNR compared to similar sequences at lower field strengths, offering promising prospects for cerebral perfusion imaging at ultrahigh field strengths.
\paragraph{}
In 2024, Kilic et al. \cite{Kilic2024} proposed an approach to mitigate B+1 inhomogeneity in multi-channel transmit arrays at 7T using unsupervised DL with CNNs. 
This approach differs from previous methodologies, which relied on supervised training, by employing unsupervised training and CNNs for multi-channel B+1 maps. 
By concatenating B+1 maps along the spatial dimension, the method enables shift-equivariant processing, which is suitable for CNNs. The training process employs a physics-driven loss function, which is designed to minimise the discrepancy between the Bloch simulation and the target magnetisation. 
The proposed method has been shown to offer superior performance compared to traditional approaches in the domain of static PTx design. The method offers enhanced speed and resilience while preventing the formation of local phase singularities.
\paragraph{}
In 2022, Plumley et al. \cite{Plumley2022} presented a new approach to address the sensitivity of head movement in adapted PTx pulses used for uniform excitation profiles at 7T. 
The researchers proposed a solution that involves reshaping pulses in real time using a DL framework. 
To train their model, simulated data is used, and conditional GANs are employed to predict B+1 distributions when there is head displacement. The accuracy of the predicted maps is evaluated against true B1 maps, considering both magnitude and phase. 
The study found that the predicted B+1 maps aligned well with the true maps and exhibited lowererror than motion-related errors in the majority of evaluations. 
Furthermore, the use of the predicted maps to redraw the pulses resulted in a significant reduction in the worst-case error of the inversion angle due to movement. 
Consequently, the proposed framework offers a solution for the real-time prediction of B+1 maps, which can be employed to redesign personalised pulses and mitigate errors related to head movement in PTx applications.
\paragraph{}
In their latest publication, Motyka et al. \cite{Motyka2024} have developed a DL approach to anticipate changes in the brain's B0 field caused by subject movement during 7T MRI scans.
The method utilises a 3D U-net trained on 7T MRI data to accurately predict B0 field alterations based on variations in head position. 
By fine-tuning the network weights to individual subjects, they achieved superior performance compared to traditional methods, enhancing spatial resolution. 
This study presents a promising solution for mitigating B0 inhomogeneities induced by subject motion, offering potential applications in retrospective or real-time correction during 7T MRI examinations.

\begin{longtable}{|p{1.5cm}|p{4.5cm}|p{4.5cm}|p{4.5cm}|}
    \caption{Summary of B0 Inhomogeneity Studies on Ultra-High-Field MRI Systems.}
    \label{tab10}\\
    \hline
    \textbf{Author} & \textbf{Purpose} & \textbf{Methods} & \textbf{Results} \\
    \hline
    \endfirsthead

    \multicolumn{4}{c}%
    {\tablename\ \thetable\ -- {Summary of B0 Inhomogeneity Studies on Ultra-High-Field MRI Systems.}} \\
    \hline
    \textbf{Author} & \textbf{Purpose} & \textbf{Methods} & \textbf{Results} \\
    \hline
    \endhead

    \multicolumn{4}{r}{{(Continues on next page)}} \\
    \endfoot

    \hline
    \endlastfoot
        Haast et al. (2021) \cite{Haast2021} & Examine the impact of MP2RAGE B1+ sensitivity on T1 reproducibility and hippocampal morphometry in 7T MRI &   The study employed the MP2RAGE sequence, which was used to examine multiple sites. &   The study demonstrated the necessity of optimising MP2RAGE sequences for neuroimaging studies and the importance of standardisation. \\ \hline
        Ma et al. (2022) \cite{Ma2022} & Mitigate B1+ artefacts in MRI without PTx hardware &   A DL model was developed to predict PTx-style images. &   The results demonstrated enhanced diffusion tensor imaging analysis, the restoration of signal dropout, and an improvement in image quality. \\ \hline
        Hunger et al. (2023) \cite{Hunger2023} & Enhance speed and precision of CEST MRI parameter mapping at 7T &   They developed a The DL approach, designated deepCEST 7T, employs a single scan prediction methodology that leverages 7T \textit{in vivo} and B1 maps. &   The scan time was reduced to 6 minutes and 42 seconds, and the ability to correct for B0 and B1 inhomogeneities was also demonstrated. \\ \hline
        Xu et al. (2023) \cite{Xu2023} & Address B0 inhomogeneity in MRI and MR spectroscopy &   The following techniques have been implemented: automated high-order shimming (autoHOS), DL-based brain extraction, and image-based high-order shimming. &   The image and spectral quality of the system have been enhanced in comparison to traditional linear and manual shimming methods. \\ \hline
        Zhao et al. (2023) \cite{Zhao2023} & Develop distortion-free 3D pCASL sequence at 7T to overcome B1/B0 inhomogeneities &   optimisation of pCASL labelling parameters, background suppression pulses, and accelerated Turbo-FLASH readout were implemented. &   The imaging technique offered high-resolution pCASL imaging, whole-brain coverage, improved SNR, and excellent test-retest repeatability. \\ \hline
        Kilic et al. (2024) \cite{Kilic2024} & Mitigate B+1 inhomogeneity in multi-channel transmit arrays at 7T &   The study employs unsupervised DL with CNN for multi-channel B+1 maps, with a physics-driven loss function. &   This approach offers superior performance compared to traditional approaches, enhanced speed and resilience, and the prevention of local phase singularities. \\ \hline
        Plumley et al. (2022) \cite{Plumley2022} & Address sensitivity of head movement in adapted PTx pulses for uniform excitation profiles at 7T &   The study used reshaping pulses using the DL framework, conditional GANs, in simulated data. &   The predicted B+1 maps matched the actual maps well, significantly reducing the worst-case error in the inversion angle due to motion. \\ \hline
        Motyka et al. (2024) \cite{Motyka2024} & Anticipate changes in brain's B0 field caused by subject movement during 7T MRI scans &   3D U-net trained on 7T MRI data: network weights fine-tuned to individual subjects. &   The methods showed superior performance compared to traditional methods, improved spatial resolution, and potential for retrospective or real-time correction. \\ \hline
\end{longtable}

\subsection{Cardiac study}
\paragraph{}
This subsection explores studies on the application of DL in cardiac MRI examinations. The studies presented here are shown in Table \ref{tab11}.
In 2021, Ankenbrand et al. \cite{Ankenbrand2021} explored the transformative potential of DL for the automatic segmentation of cardiac cine MRI images. Employing transfer learning techniques, the researchers adapted existing models to the realm of 7T MRI. 
By leveraging a publicly available segmentation model to annotate a dataset and subsequently training a neural network to segment the left ventricle and myocardium, the study revealed that transfer learning significantly boosted model performance, even when applied to a reduced dataset. 
Furthermore, this research provided practical guidelines and resources to facilitate similar endeavours in the future.
\paragraph{}
Van der Brick et al. \cite{vanderbrinck} presented the principles and design of the ZOOM@SVDs research, which aims to determine measures of cerebral small vessel dysfunction using 7T MRI as a new method for small vessel disease (SVD). 
The study includes an observational cohort of patients with cerebral autosomal dominant arteriopathy with subcortical infarcts and leukoencephalopathy, sporadic SVDs, and healthy controls. 
The study uses a 7T brain MRI to evaluate features of small vessel function such as its reactivity, cerebral punching artery flow and pulsatility.  
The evaluation included clinical and neuropsychological assessments, as well as conventional 3T MRI to evaluate conventional imaging markers of SVD. 
In conclusion, by comparing measures of small vessel dysfunction between patients and controls and correlating them with clinical and MRI manifestations of SVD, ZOOM@SVDs identified novel markers of cerebral small vessel function.
\paragraph{}
In a complementary study, Garcia-garcia et al. \cite{Garciagarcia} introduced Vessel Distance Mapping (VDM) to assess blood supply and its relationship to cognition. 
The researchers aimed to overcome previous controversies regarding the connection between cerebral blood flow and cognitive function. 
VDM provides a quantitative approach to evaluating vascular patterns about surrounding structures, extending the previous binary classification into a continuous spectrum. 
Using 7T MR angiographic imaging, the researchers manually segmented hippocampal vessels in older adults with and without cerebral SVD. 
VDM metrics, which indicate vessel distances, were found to correlate with cognitive results in individuals with vascular pathology but not in healthy controls. 
The study highlights that both vessel pattern and vessel density contribute to cognitive resilience. 
VDM provides a reliable method for vascular mapping, enabling the investigation of various clinical research questions related to cerebral blood flow and cognition.
\paragraph{}
In 2023, Chowdhury et al. \cite{Chowdhury2023} conducted an investigation into the detection of R peaks in ECG waveforms that had been distorted by the magnetohydrodynamic effect during MRI scans, with a particular focus on 3T and 7T magnetic fields.
The researchers proposed the Self-Attention MHDNet model, which was designed to accurately identify R-peaks from distorted ECG signals. The model demonstrated high recall and precision rates, with 99.83\% and 99.68\% in 3T settings, and 99.87\% and 99.78\% in 7T settings, respectively. 
This improved detection capability can enhance the precision of functional cardiovascular MRI by ensuring the reliable triggering of the pulse.
The study demonstrates the potential of integrating machine learning techniques to mitigate ECG signal distortions in the MRI environment, thereby facilitating enhanced patient monitoring and diagnosis.
\paragraph{}
In 2024, Kolmann et al. \cite{Kollmann2024} sought to improve the reproducibility of image segmentation for myocardial tissue analysis in a large animal model of myocardial infarction using CMR imaging at 7T. 
The researchers discovered that traditional automated segmentation approaches, designed for human imaging at clinical field strengths, encountered difficulties when applied to preclinical data and ultra-high field strengths. 
To address this issue, the researchers re-trained a DL model using their preclinical 7T data and tested it on a separate dataset. 
The DL segmentation showed excellent agreement with manual segmentation, as evidenced by a high Pearson correlation coefficient, intraclass correlation coefficient and low coefficient of variability for ejection fraction. The Dice scores for left ventricular and myocardial segmentation were also high. 
This approach shows great promise for improving the analysis of cardiac function, particularly in large animal models of myocardial infarction.

\begin{longtable}{|p{1.5cm}|p{4.5cm}|p{4.5cm}|p{4.5cm}|}
    \caption{Summary of Cardiac Studies on Ultra-High-Field MRI Systems.}
    \label{tab11}\\
    \hline
    \textbf{Author} & \textbf{Purpose} & \textbf{Methods} & \textbf{Results} \\
    \hline
    \endfirsthead

    \multicolumn{4}{c}%
    {\tablename\ \thetable\ -- {Summary of Cardiac Studies on Ultra-High-Field MRI Systems.}} \\
    \hline
    \textbf{Author} & \textbf{Purpose} & \textbf{Methods} & \textbf{Results} \\
    \hline
    \endhead

    \multicolumn{4}{r}{{(Continues on next page)}} \\
    \endfoot

    \hline
    \endlastfoot
        Ankenbrand et al. (2021) \cite{Ankenbrand2021} & Automatic segmentation of cardiac cine MRI images using DL &   Use of transfer learning to adapt models to 7T MRI. &   The proposed methods showed improved segmentation performance with a reduced dataset, providing practical guidelines for similar projects. \\ \hline
        van der Brick et al. \cite{vanderbrinck} & Establish measures of cerebral small vessel dysfunction using 7T MRI &   The technique was tested in a prospective observational cohort study with subjects with CADASIL, sporadic SVDs, and healthy controls. &   Novel markers of small vessel function, providing insights into the pathophysiology of SVD. \\ \hline
        Garcia-garcia et al. \cite{Garciagarcia} & Assess blood supply and its relation to cognition using VDM &   The VDM was used in conjunction with a 7T MR angiographic imaging system. &   The results showed a correlation between vessel distances and cognitive outcomes in individuals with vascular pathology. \\ \hline
        Chowdhury et al. (2023) \cite{Chowdhury2023} & Detect R-peaks in ECG waveforms distorted by magnetohydrodynamic effect during MRI &   The researchers developed a Self-Attention MHDNet model. &   The proposed model showed high recall and precision rates (99.83\%/99.68\% at 3T, 99.87\%/99.78\% at 7T). \\ \hline
        Kollmann et al. (2024) \cite{Kollmann2024} & Enhance reproducibility of image segmentation for myocardial tissue analysis in a large animal model of myocardial infarction &   The researchers re-trained a DL model tested on preclinical 7T data. &   It demonstrated excellent agreement with manual segmentation, high Pearson and ICC, and low variability. \\ \hline
\end{longtable}

\subsection{DL study}
\paragraph{}
This subsection explores studies on the application of general DL, to generate and reconstruct reliable MRI images. The studies presented here are shown in Table \ref{tab12}.
In 2018, Nie et al. \cite{Nie2018} proposed a technique to generate medical images using an FCN from a source image. 
The researchers used adversarial learning strategies to improve the realism of the generated images.  In addition, a loss function based on the image gradient difference was implemented to prevent the generation of blurred target images. 
The method proposed in this study demonstrated both accuracy and robustness in experiments performed on a variety of datasets. 
Consequently, these tasks, such as the synthesis of CT images from MRI scans and the generation of 7T MRI images from 3T MRI scans, demonstrated superior performance compared to existing approaches.
\paragraph{}
In a study addressing the issue of accessibility to expensive 7T MRI scanners, Qu et al. \cite{Qu2020} conducted a study. 
A DL network was developed to synthesise high-quality 7T MRI images from their 3T counterparts. 
The network incorporates information from both the spatial and wavelet domains, including a novel wavelet-based affine transformation layer. The network's use of the wavelet transform for effective multi-scale reconstruction improves tissue contrast and anatomical detail in synthesised 7T images. 
The results show the effectiveness of the proposed method, which outperforms existing techniques and offers a promising solution to the limited accessibility of 7T MRI scanners.
\paragraph{}
In 2024, Cabini et al. \cite{Cabini2024} presented a DL model to enhance the reconstruction of T1 and T2 maps obtained through MRF on a 7T MRI machine. 
The implementation of a hyperparameter optimisation strategy enabled the model to achieve enhanced reconstruction accuracy and speed for preclinical investigations. 
In comparison to traditional methodologies, the proposed DL approach demonstrated a reduction in reconstruction errors. This was achieved through the optimisation of neural network architectures, model structures, and supervised learning algorithms. 
Furthermore, it enhanced computational efficiency, rendering it suitable for real-time applications. 
The results demonstrated that the DL method outperformed dictionary-based methods, resulting in a reduction of the mean percentage relative error.
These findings indicate the potential of the DL model to significantly advance MRI reconstruction techniques.
\paragraph{}
Eidex et al. \cite{Eidex2024} proposed a hybrid CNN-transformer model to generate high-resolution 7T ADC maps from multimodal 3T MRI data. This addresses the limitations of 7T MRI, such as limited availability and higher costs.
Their Vision CNN-Transformer model, which combines Vision Transformer blocks for global context and convolutional layers for fine details, was validated using the Human Connectome Project Young Adult dataset. 
The model was trained on 130 cases, validated on 20, and tested on 21. 
The Vision CNN-Transformer model achieved a PSNR of 27.0 ± 0.9 dB, SSIM of 0.945 ± 0.010, and MSE of 2.0E-3 ± 0.4E-3, outperforming other methods and reducing training time by 50\%. 
The findings demonstrated that the Vision CNN-Transformer model was capable of effectively synthesising 7T quality ADC maps from 3T data, which could potentially enhance disease diagnosis and intervention.

\begin{longtable}{|p{1.5cm}|p{4.5cm}|p{4.5cm}|p{4.5cm}|}
    \caption{Summary of general DL Studies on Ultra-High-Field MRI Systems.}
    \label{tab12}\\
    \hline
    \textbf{Author} & \textbf{Purpose} & \textbf{Methods} & \textbf{Results} \\
    \hline
    \endfirsthead

    \multicolumn{4}{c}%
    {\tablename\ \thetable\ -- {Summary of DL Studies on Ultra-High-Field MRI Systems.}} \\
    \hline
    \textbf{Author} & \textbf{Purpose} & \textbf{Methods} & \textbf{Results} \\
    \hline
    \endhead

    \multicolumn{4}{r}{{(Continues on next page)}} \\
    \endfoot

    \hline
    \endlastfoot
        Nie et al. (2018) \cite{Nie2018} & Generate medical images using a fully convolutional network (FCN) from a source image &   The methodologies employed encompass adversarial learning and the image gradient difference loss function. &   The results demonstrated the accurate and robust synthesis of CT from MRI and 7T MRI from 3T MRI, with superior performance compared to existing approaches. \\ \hline
        Qu et al. (2019) \cite{Qu2020} & Synthesise high-quality 7T MRI images from 3T MRI images &   The researchers developed a DL network comprising a wavelet-based affine transformation layer. &   The technique offers enhanced tissue contrast and anatomical detail, as well as effective multi-scale reconstruction, which outperforms current techniques. \\ \hline
        Cabini et al. (2024) \cite{Cabini2024} & Enhance reconstruction of T1 and T2 maps from MRF on 7T MRI &   A DL model was developed to optimise the hyperparameters. &   A significant reduction in reconstruction errors, accompanied by enhanced computational efficiency, renders the method suitable for real-time applications. \\ \hline
        Eidex et al. (2024) \cite{Eidex2024} & Generate high-resolution 7T ADC maps from multimodal 3T MRI data &   Develop the Hybrid CNN-transformer model, a Vision CNN-Transformer. &   The method demonstrated high PSNR and SSIM values, with a 50\% reduction in training time;
        
          It was able to successfully synthesise 7T-quality ADC maps from 3T data. \\ \hline
\end{longtable}

\paragraph{}
\section{Discussion}
\paragraph{}
The rise of 0.55T and 7T MRI technologies reveals a transformative leap in medical imaging, with each offering distinctive benefits and challenges. The convergence of these innovative systems with DL techniques unlocks the potential to improve diagnostic imaging practices. By enhancing image quality, accelerating acquisition times, and elevating diagnostic accuracy, this synergistic integration holds the power to redefine the boundaries of what is achievable.
\paragraph{}
This review has delved into the developments in 0.55T and 7T MRI, highlighting the transformative impact of DL integration.
The progress in 0.55T low-field MRI has shown potential for increased accessibility and cost-effectiveness, while 7T MRI still sets new standards for image resolution and functional imaging capabilities.
\paragraph{}
Historically, 0.55T MRIs have been limited by lower SNR and CNR compared to higher-field magnets, potentially compromising image quality and diagnostic efficacy. 
However, recent pioneering studies have unveiled the ability of DL techniques to fill this gap. The Siemens Healthineer's High-V scanner, powered by the Deep Resolve DL model, exemplifies this potential. 
Furthermore, the studies presented in this review underscore the feasibility of integrating DL into low-field MRI for optimised imaging protocols across various body regions. 
Studies such as those by Rusche et al. \cite{Rusche2022_confort}, Nayak et al. \cite{Nayak2023}, and Kim et al. \cite{Kim2024} have demonstrated the potential of 0.55T MRI in enhancing patient comfort, body composition profiling, and accelerating MR cholangiopancreatography acquisitions through DL-based reconstruction techniques. 
\paragraph{}
These findings collectively indicate that while DL addresses the inherent limitations of lower magnetic field strengths, such as lower SNR and CNR, 0.55T MRI technology still has restrictions in terms of spatial resolution and functional/spectroscopic imaging capabilities, as well as requiring longer acquisition times. However, the integration of DL opens up new frontiers for clinical practice, allowing high-quality images to be obtained at affordable costs. 0.55T MRI systems offer a number of advantages, including lower acquisition and operating costs, a more compact size and weight, making installation easier, less need for shielding against intense magnetic fields, and suitability for most routine clinical protocols. 
The use of DL techniques enables the acquisition of high-quality images, despite the inherent limitations of low-field systems in terms of SNR.
\paragraph{}
7T MRI systems offer high spatial resolution and SNR, positioning them as essential tools for complex anatomical and functional imaging. 
The integration of DL techniques has expanded the potential of 7T MRI in medical imaging. This synergy has accelerated the development of DL algorithms capable of managing high data rates, attenuating artefacts and optimising image reconstruction. 
These advancements have solidified the clinical value of 7T MRI.
Numerous case studies have demonstrated DL's remarkable ability to speed up acquisitions while preserving quality and reducing digitisation times, thus improving diagnostic imaging.
Studies such as those by Ramadass et al. \cite{Ramadass2023}, Tsuji et al. \cite{Tsuji2023}, Shamir et al. \cite{Shamir2019} and Motyka et al. \cite{Motyka2024} have highlighted the superior spatial resolution, advancements in whole brain segmentation, and the reduction of acquisition time through the utilisation of a residual dense network. These studies have also demonstrated improvements in visualising fine anatomical structures, such as the subthalamic nucleus in Parkinson's patients, and have showcased the advanced functional imaging capabilities provided by 7T MRI systems. 
These studies underline the fundamental role of DL in optimising image quality, speeding up workflows and enabling advanced imaging applications previously unfeasible in ultra-high field MRI. 
The integration of DL algorithms with 7T MRI not only improves diagnostic accuracy, but also expands the scope of high-resolution clinical and research applications.
\paragraph{}
The findings demonstrate that 7T MRI has the following advantages: high SNR and spatial resolution, excellent visualisation of fine anatomical structures and details, and superior imaging capabilities. DL enables the acquisition of high-quality images at ultra-fast speeds, making it an ideal choice for research and advanced high-resolution clinical applications. 
However, the high acquisition and operating costs associated with this technology require special installations with shielding against intense fields. Furthermore, it has been demonstrated that there is a greater potential for image artefacts and distortions, with high risks of energy deposition in patients and heating, which makes it contraindicated for some implants and medical devices.
The articles reviewed indicate that higher magnetic fields are recommended, particularly for applications requiring extremely high spatial resolution and advanced functional visualisation imaging.
\paragraph{}
The main differences in the use of 0.55T and 7T MRI with DL are as follows:
\paragraph{}
\textbf{0.55T MRI:}
\begin{itemize}
    \item Generally smaller data sets with lower SNR, which makes training complex DL models more challenging.
    \item DL is widely used for image enhancement, noise removal and reconstruction/super-resolution from undersampled data.
    \item Transfer learning from models trained on high-field data helps compensate for the limitations of low-field data.
\end{itemize}
\paragraph{}
\textbf{7T MRI:}
\begin{itemize}
    \item Large, high-quality, high SNR/CNR datasets suitable for training complex, data-intensive DL models.
    \item DL is also used to correct artefacts, distortions and acquisition accelerations.
    \item DL is essential for processing and visualising the multidimensional data of high-resolution functional/spectral images.
\end{itemize}
\paragraph{}
The main focus is on using DL models to help to achieve very short scan times while preserving maximum image quality. The focus is on harnessing the wealth of data to advance high-resolution clinical/research applications. 
While 0.55T MRI primarily uses DL to overcome data limitations and achieve acceptable image quality, inherent to its limitations of lower SNR/CNR. 
7T systems employ DL to extract insights from rich data, accelerate workflows and enable previously unfeasible advanced imaging applications. The availability and quality of data significantly influence the scope and nature of DL applications.

\subsection{Current Challenges and Future Directions}
\paragraph{}
Although the integration of DL into MRI is very promising, it is not without its challenges. The demand for substantial computational resources, the complexity of training algorithms, and the potential for algorithmic bias represent significant considerations that must be addressed \cite{Heckel2024}. 
Additionally, the high field strengths inherent to 7T MRI can aggravate side effects and image artefacts, requiring the implementation of advanced management and correction techniques.
For 0.55T MRI, the primary challenge lies in achieving image quality comparable to higher field strengths. Although DL techniques have shown the potential to fill this gap, further validation and optimisation are needed to guarantee clinical reliability \cite{Li2024}.
Moreover, the complexity of DL models and the need for extensively annotated datasets present obstacles for both 0.55T and 7T MRI. 
\paragraph{}
Looking ahead, future research efforts should focus on optimising DL algorithms adapted to specific MRI modalities, addressing computational challenges and improving the robustness and reliability of DL-based imaging techniques.
Furthermore, the clinical validation of these integrated technologies must be extended to enable widespread adoption and implementation in routine clinical practice \cite{Li2024}, \cite{Shimron2023}. To advance the integration of DL into MRI, future research should prioritise the following areas:

\begin{itemize}
    \item \textbf{Data Sharing:} Establishing large-scale, multi-centre collaborations to create comprehensive, annotated datasets will be crucial for training robust DL models that can be generalised effectively.
    \item \textbf{Model generalisation:} Developing and validating DL models that can be generalised across diverse populations and clinical contexts, ensuring widespread applicability and equitable access to these cutting-edge imaging solutions.
    \item \textbf{Cost-Benefit Analysis:} Conducting detailed cost-benefit analyses to evaluate the economic impact of implementing advanced DL techniques in resource-constrained settings, particularly for 0.55T MRI, which holds potential for expanding access to high-quality imaging in poorly served regions \cite{Shimron2023}.
\end{itemize}

\section{Conclusion}
\paragraph{}
The integration of deep learning techniques with 0.55T and 7T MRI systems marks a transformative milestone in the field of medical imaging. 
This convergence reveals an ability to visualise complex anatomical structures and functional processes with remarkable clarity. 
The 0.55T MRI represents a more cost-effective solution for clinical routine, while the 7T systems, although more expensive, are powerful tools for research and cases that require images of extremely high resolution and quality. 
Innovative approaches like Siemen's High-V scanner with Deep Resolve have demonstrated the potential to reduce noise and elevate image quality in low-field MRI, rendering it comparable to higher-field systems. Such advancements promise reinforced diagnostic precision while expanding patient accessibility through more cost-effective imaging protocols across diverse body regions.
The studies reviewed shed light on the vast diagnostic usefulness of ultra-high-field MRI, from revealing details of brain functionality to assessing the structural integrity associated with neurological conditions.
Advances in DL-guided segmentation and reconstruction methods have enabled improvements in image accuracy, speed and fidelity, extending the clinical impact of ultra-high-field MRI.
Looking to the future, ongoing research synergising DL with MRI will be important for the development of increasingly sophisticated applications to redefine diagnostic capabilities and patient care standards.  
\newline

\textbf{Acknowledgments}

This work was supported by the REACT-EU project KITE (Plattform für KI-Translation Essen, EFRE-0801977, \url{https://kite.ikim.nrw/}), FWF enFaced 2.0 (KLI 1044, \url{https://enfaced2.ikim.nrw/}), AutoImplant (\url{https://autoimplant.ikim.nrw/}) and “NUM 2.0” (FKZ: 01KX2121). Behrus Puladi was funded by the Medical Faculty of RWTH Aachen University as part of the Clinician Scientist Program. André Ferreira was funded by the Fundação para a Ciência e Tecnologia (FCT) Portugal with the grant 2022.11928.BD. Furthermore, we acknowledge the Center for Virtual and Extended Reality in Medicine (ZvRM, \url{https://zvrm.ume.de/}) of the University Hospital Essen.

\newpage
\bibliographystyle{unsrt}  
\bibliography{references}

\end{document}